\documentclass[a4paper,fleqn,review]{cas-dc}

\usepackage[authoryear]{natbib}
\bibpunct{(}{)}{;}{a}{,}{,}
\usepackage{hyperref}
\usepackage{multirow}
\usepackage{arydshln}
\usepackage{rotating, float, caption}
\usepackage{setspace}   %Allows double spacing with the \doublespacing command
\usepackage[normalem]{ulem}

\def\tsc#1{\csdef{#1}{\textsc{\lowercase{#1}}\xspace}}
\tsc{WGM}
\tsc{QE}
\tsc{EP}
\tsc{PMS}
\tsc{BEC}
\tsc{DE}
\begin{document}
\let\WriteBookmarks\relax
\def\floatpagepagefraction{1}
\def\textpagefraction{.001}
\shorttitle{Earth co-orbitals properties}
\shortauthors{G. Borisov et~al.}

\title [mode = title]{Physical and dynamical properties of selected Earth co-orbital asteroids}                      
\tnotemark[1]

\tnotetext[1]{Partially based on data collected with the 2-m\,RCC telescope at the Bulgarian National Astronomical Observatory - Rozhen}

%\tnotetext[2]{The second title footnote which is a longer text matter
%   to fill through the whole text width and overflow into
%   another line in the footnotes area of the first page.}

\author[1,2]{Galin B. Borisov}[orcid=0000-0002-4516-459X]
\cormark[1]
\fnmark[1]
\ead{Galin.Borisov@Armagh.ac.uk}
%\ead{gborisov@astro.bas.bg}
%\ead[url]{www.cvr.cc, cvr@sayahna.org}

%\credit{Conceptualization of this study, Methodology, Software}

\address[1]{Armagh Observatory and Planetarium, College Hill, Armagh BT61 9DG, Northern Ireland, United Kingdom}

\author[1]{Apostolos A. Christou}

%\fnmark[2]
%\ead{cvr3@sayahna.org}
%\ead[URL]{www.sayahna.org}

%\credit{Data curation, Writing - Original draft preparation}

\address[2]{Institute of Astronomy with NAO, Bulgarian Academy of Sciences, 72 Tsarigradsko Chauss\'ee Blvd, BG-1784, Sofia, Bulgaria}

\author[3]{Gordana Apostolovska}

\address[3]{Institute of Physics, Faculty of Natural Sciences and Mathematics,  Ss. Cyril and Methodius University in Skopje, Arhimedova 3, 1000 Skopje, FYR of Macedonia}

%\author[1,3]
%{Rishi T.}
%\cormark[2]
%\fnmark[1,3]
%\ead{rishi@stmdocs.in}
%\ead[URL]{www.stmdocs.in}

%\address[3]{STM Document Engineering Pvt Ltd., Mepukada,
%    Malayinkil, Trivandrum 695571, India}

\cortext[cor1]{Corresponding author}
%\cortext[cor2]{Principal corresponding author}
%\fntext[fn1]{This is the first author footnote. but is common to third
%  author as well.}
%\fntext[fn2]{Another author footnote, this is a very long footnote and
%  it should be a really long footnote. But this footnote is not yet
%  sufficiently long enough to make two lines of footnote text.}

%\nonumnote{This note has no numbers. In this work we demonstrate $a_b$
%  the formation Y\_1 of a new type of polariton on the interface
%  between a cuprous oxide slab and a polystyrene micro-sphere placed
%  on the slab.
%  }
% \doublespacing

\begin{abstract}
We present our investigations of the physical and dynamical properties of selected Earth co-orbital asteroids. The photometric optical light curves as well as rotation periods and  pole solutions for a sample of four Earth co-orbital asteroids, namely (138175) 2000 EE104 (P=13.9476$\pm$0.0051\,hrs), (418849) 2008 WM64 (P=2.4077$\pm$0.0001\,hrs), 2016 CA138 (P=5.3137$\pm$0.0016\,hrs) and 2017 SL16 (P=0.3188$\pm$0.0053\,hrs), are determined or improved and presented in this work. For this investigation, we combine observations carried out at the Bulgarian National Astronomical Observatory - Rozhen using the FoReRo2 instrument attached to the 2mRCC telescope as well as sparse data from AstDys2 database. Parallel to the rotational properties we did numerical dynamical simulations to investigate the orbital stability of those objects and to find out if there is a relation with their rotational properties. Our results show that the orbit stability is affected by the orbit itself and mainly its eccentricity and inclination, which are responsible for the close encounters with other Solar system planets.
We cannot make a definitive conclusion about the relation between orbit stability and the rotational state of the asteroids, so we need further investigations and observations in order to prove or disprove it.
%\baselineskip=2em
% \doublespacing
\end{abstract}

%\begin{graphicalabstract}
%\includegraphics{figs/grabs.pdf}
%\end{graphicalabstract}

%\begin{highlights}
%\item New photometric optical light curves of four Earth co-orbital asteroids are obtained
%\item Rotation periods and pole solutions of those objects are determined
%\item Lightcurve inversion was applied to improve the period determination
%\item Co-orbital asteroids appear to have rotation rates consistent to NEAs of similar size
%\item Dynamical stability is correlated with the orbit, but not with the resonant state 
%\end{highlights}

\maketitle

% \doublespacing

\section{Introduction}
Asteroids with an average heliocentric distance of 1\,au - also called the Earth co-orbital asteroids - present a special challenge to Earth-based surveys. Because of the very slow net relative motion  with respect to the Earth from one orbital revolution to the next, they usually remain far from our planet but close to the Sun's projection to the sky for many years. This effect leads to a much lower observational completeness for these types of objects than for other near-Earth asteroids (NEAs) \citep{Tricarico2017}. When a co-orbital object is discovered, it typically offers a few brief annually-recurring apparitions when it is bright enough to allow physical characterisation. Afterwards, the accumulated Keplerian drift due to the slight difference in orbital frequency between Earth and the asteroid will place it out of reach of observational scrutiny for many decades hence.

 The dynamical behaviour of Earth coorbitals continues to be an area of active research \citep{DiRuzza.et.al2023,QiQiao2022} ever since the discovery of the first member of this class, (3753) Cruithne \citep{Wiegert.et.al1997} .
Co-orbital asteroids are generally thought to have more stable orbits against encounters with the planets, where ``stable'' is here taken to mean that the semimajor axis $a$, eccentricity $e$ and inclination $I$ diffuse slowly over time or not at all. This is partly due to relatively long intervals between planetary conjunctions but also because, even for those asteroids that approach the planet, the resonant condition  generally yields  only shallow encounters where the resulting orbit change is both small and deterministic,  leading to so-called {\it transitions} between coorbital modes  \citep{Namouni1999,Christou2000}.

The principal driver of rotational and physical changes in main belt asteroids smaller than $\sim$6\,km is the non-gravitational YORP effect, causing significant changes to the spin state over a timescale that is size-dependent but typically <$10^{7}$\,yr \citep{Rubincam2000,Jacobson2014}. In the terrestrial planet region, YORP competes with the torques and tides exerted across the asteroid during close planetary encounters \citep{Scheeres.et.al2004,WalshRichardson2006,Walsh.et.al2008}. It stands therefore to reason that, if close encounters are important to the physical and rotational evolution of NEAs, the properties of Earth co-orbitals might differ from those of other NEAs as the resonant condition offers some degree of protection from the closest encounters.  

In this paper, we report on rotational state of a sample of Earth co-orbital asteroids. Of the four co-orbital NEAs in our sample, two were the subject of previous work \citep{Borisov2021} while the remaining two are investigated here for the first time. We generate rotational pole solutions for all four asteroids. In parallel, we carry out orbit simulations of these four objects to establish which are actually locked in the 1:1 mean motion resonance with the Earth and quantify their long-term orbital stability. Afterwards, we use this information to verify, in the first instance, whether the co-orbital state promotes long-term stability of the orbit and, secondly, to compare the ensemble rotational data and orbital data between different populations: resonant as well as non-resonant asteroids at 1\,au and NEAs. 

The paper is organised as follows: in the next section, we present the photometric observations utilised in the   rotational state modelling. Section\,3 describes the data reduction with emphasis on the different approaches used to estimate the rotational state, while Section\,4 presents our results separately for the four asteroids and compares them to the NEA population. Section\,5 is dedicated to the numerical simulations to investigate the orbit dynamical properties and search for relationships to the rotational state. Finally, Section\,6 outlines our conclusions.

\section{Observations}
Photometric observations of selected Earth co-orbital asteroids were carried out from the Bulgarian National Astronomical Observatory - Rozhen, using  the Two-channel Focal Reducer Rozhen or “FoReRo2” instrument attached to the 2-m\,RCC telescope. The asteroid targets and their observational circumstances are presented in Table~\ref{TBL:Obs}.
\begin{table*}[width=1.0\textwidth,cols=7,pos=h!]
\caption{Observing circumstances and aspect data}\label{TBL:Obs}
\begin{minipage}{1.0\textwidth}\centering
\begin{tabular*}{\tblwidth}{@{} RLLRRRL@{} }
\toprule
Number & Designation & yyyy mm dd & Phase($^{\circ}$)\footnote{The phase angle is given for the first date.} & L$_{\rm PAB}$\footnote{The approximate phase angle bisector longitude at mid-date range} & B$_{\rm PAB}$\footnote{The approximate phase angle bisector latitude at mid-date range} & Grp\footnote{The asteroid family/group (APO-Apollo, ATE-Aten)} \\
\midrule
(418849) & 2008 WM64 & 2017 12 25 & 36.9 & 96 & 19 & APO \\
(138175) & 2000 EE104 & 2018 11 09 & 66.0 & 100 & 8 & APO \\
{--- \raisebox{-0.5ex}{"} ---} & {--- \raisebox{-0.5ex}{"} ---} & 2019 01 01 & 19.3 & 108 & 14 & APO \\
{--- \raisebox{-0.5ex}{"} ---} & {--- \raisebox{-0.5ex}{"} ---} & 2020 01 02 & 17.9 & 104 & 14 & APO \\
%522684 & 2016 JP & 2020 04/25,04/27 & 53.3 & 240 & 21 & ATE \\
 & 2017 SL16 & 2020 09 22 & 30.0 & 14 & 6 & ATE \\
 & 2016 CA138 & 2020 02 17\&18 & 20.3 & 158 & -7 & ATE \\
\bottomrule
\end{tabular*}
\end{minipage}
\end{table*}

\begin{table}[width=1.0\columnwidth,cols=5,pos=h!]
\caption{Sparse data used}\label{TBL:Sparse}
\begin{minipage}{1.0\columnwidth}\centering
\begin{tabular*}{\tblwidth}{@{} RLLCC@{} }
\toprule
Number & Designation & yyyy mm dd & Filter & Obs. Code\footnote{Minor Planet Center (MPC) observatory codes} \\
\midrule
(418849) & 2008 WM64  & 2019 01 07 & c & T08\footnote{T08 -- Asteroid Terrestrial-impact Last Alert System (ATLAS-MLO; at Mauna Loa Observatory)} \\
         &            & 2019 01 09 & c & T05\footnote{T05 -- Asteroid Terrestrial-impact Last Alert System (ATLAS-HKO; at Haleakala Observatory)} \\
\cdashline{3-5}
         &            & 2018 12 19 & o & T08 \\
         &            & 2018 12 21 & o & T05 \\
         &            & 2018 12 24 & o & T08 \\
         &            & 2018 12 28 & o & T05 \\
         &            & 2020 01 02 & o & T08 \\
         &            & 2020 12 20 & o & T08 \\
         &            & 2020 12 22 & o & T05 \\
         &            & 2020 12 29 & o & T08 \\
         &            & 2021 01 03 & o & T08 \\
\hline
(138175) & 2000 EE104 & 2019 01 09 & c & T05 \\
         &            & 2019 12 27 & c & T05 \\
         &            & 2020 12/11 & c & T08 \\
\cdashline{3-5}
         &            & 2018 12 28 & o & T05 \\
         &            & 2019 12 08 & o & T08 \\
         &            & 2020 01 02 & o & T08 \\
         &            & 2020 12 09 & o & T05 \\
\cdashline{3-5}
         &            & 2020 10 19 & G & G96\footnote{G96 -- Mount Lemmon Survey} \\
         &            & 2020 11 12 & G & 703\footnote{703 -- Catalina Sky Survey, Tucson} \\
         &            & 2020 11 24 & G & 703 \\
         &            & 2020 11 29 & G & G96 \\
         &            & 2020 12 19 & G & G96 \\
         &            & 2020 12 20 & G & 703 \\
         &            & 2020 12 26 & G & 703 \\
         &            & 2021 01 03 & G & G96 \\
         &            & 2021 01 05 & G & 703 \\
\hline
         & 2017 SL16  & 2020 09 23 & G & I52\footnote{I52 -- Mount Lemmon Observatory (CHECK: Mount Lemmon Survey) of the Steward Observatory} \\
\hline
         & 2016 CA138 & 2019 02 15 & o & T08 \\
\cdashline{3-5}
         &            & 2020 02 16 & g & I41\footnote{I41 -- Zwicky Transient Facility (ZTF) and its predecessor Palomar Transient Factory (PTF) at Palomar Observatory} \\
\cdashline{3-5}
         &            & 2020 02 16 & R & I41 \\
\bottomrule
\end{tabular*}
\end{minipage}
\end{table}

\begin{figure*}
	\centering
	\includegraphics[width=0.5\columnwidth]{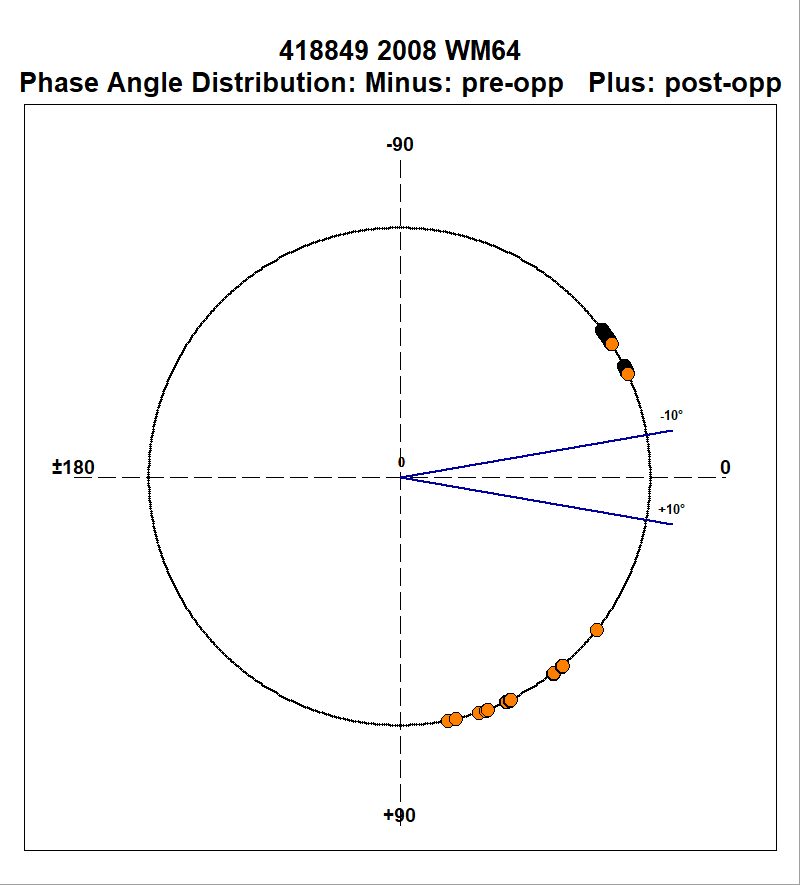}
    \includegraphics[width=0.5\columnwidth]{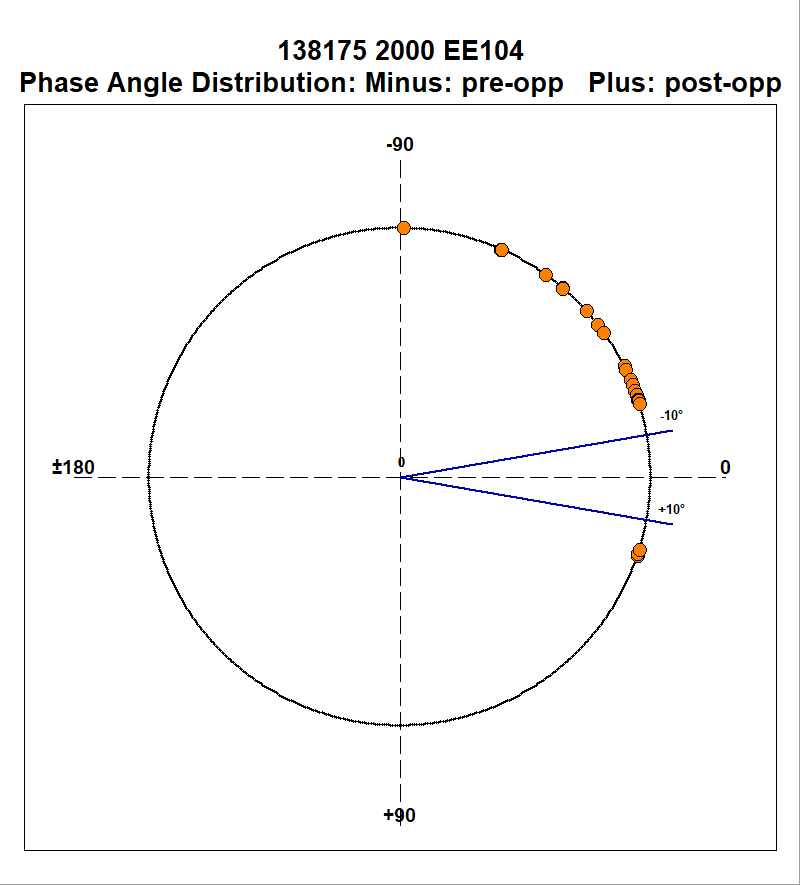}
	\includegraphics[width=0.5\columnwidth]{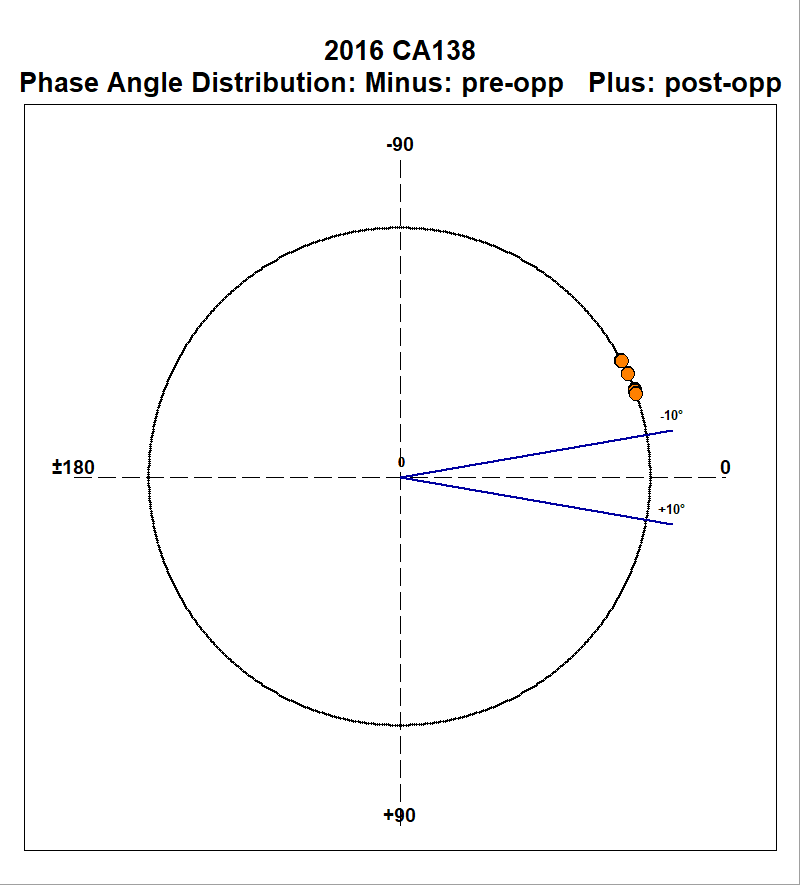}
    \includegraphics[width=0.5\columnwidth]{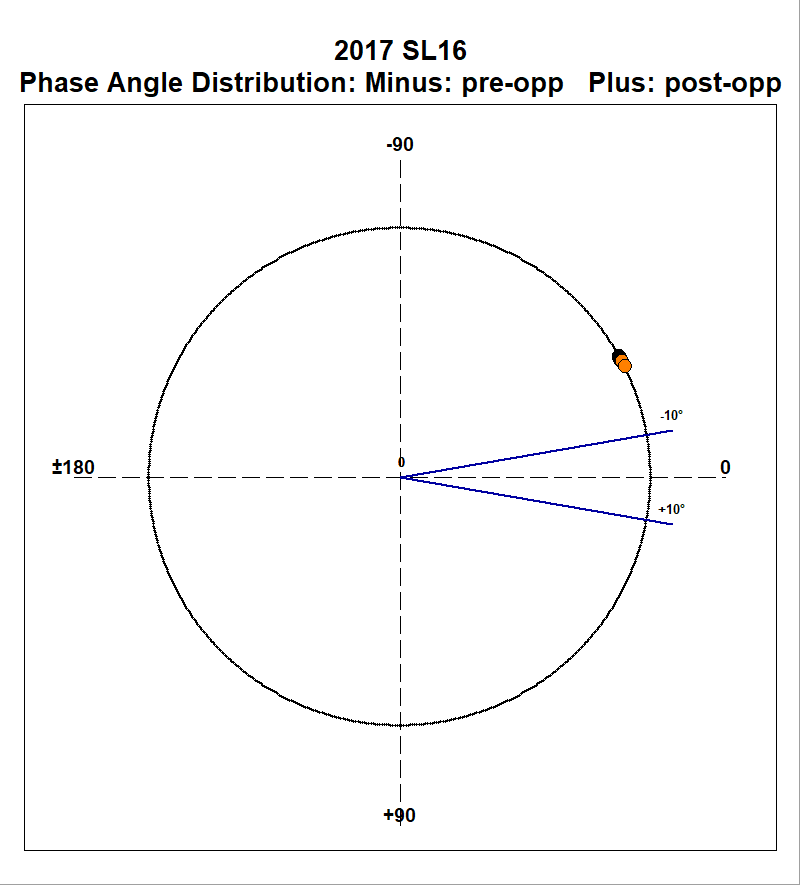}
	
	\includegraphics[width=0.5\columnwidth]{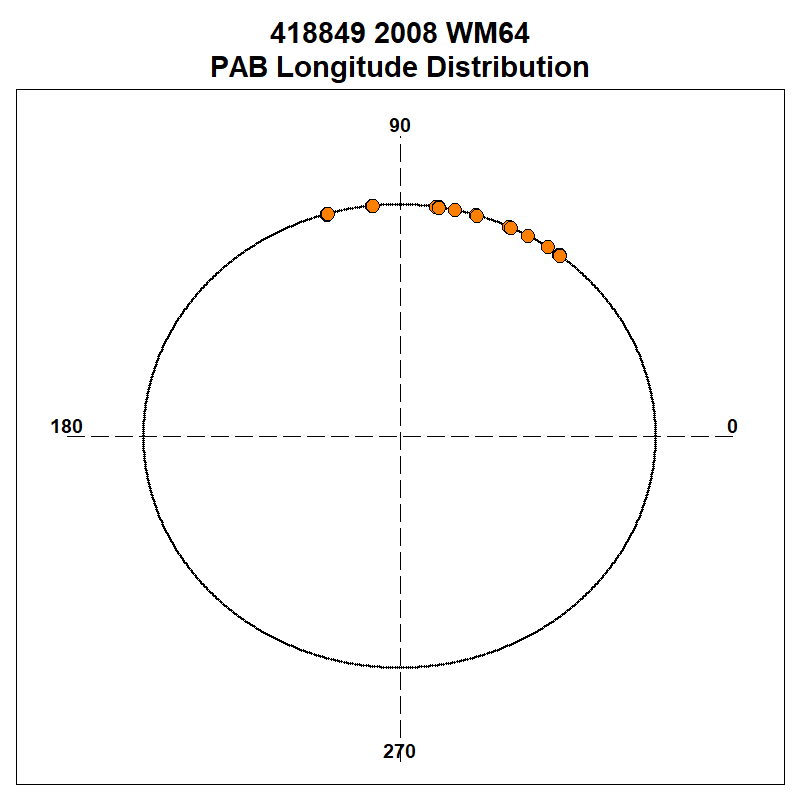}
    \includegraphics[width=0.5\columnwidth]{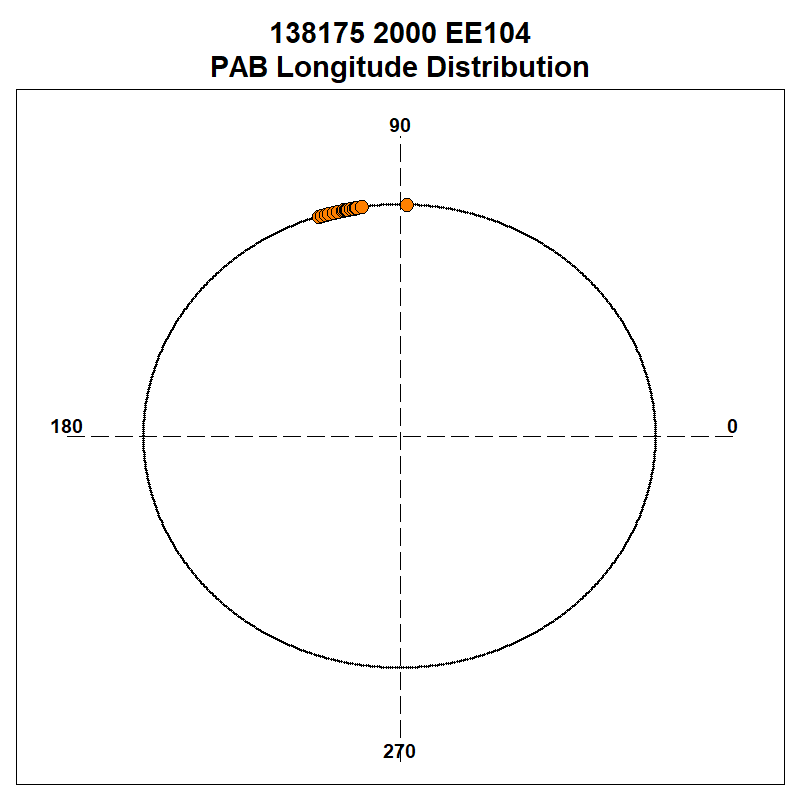}
	\includegraphics[width=0.5\columnwidth]{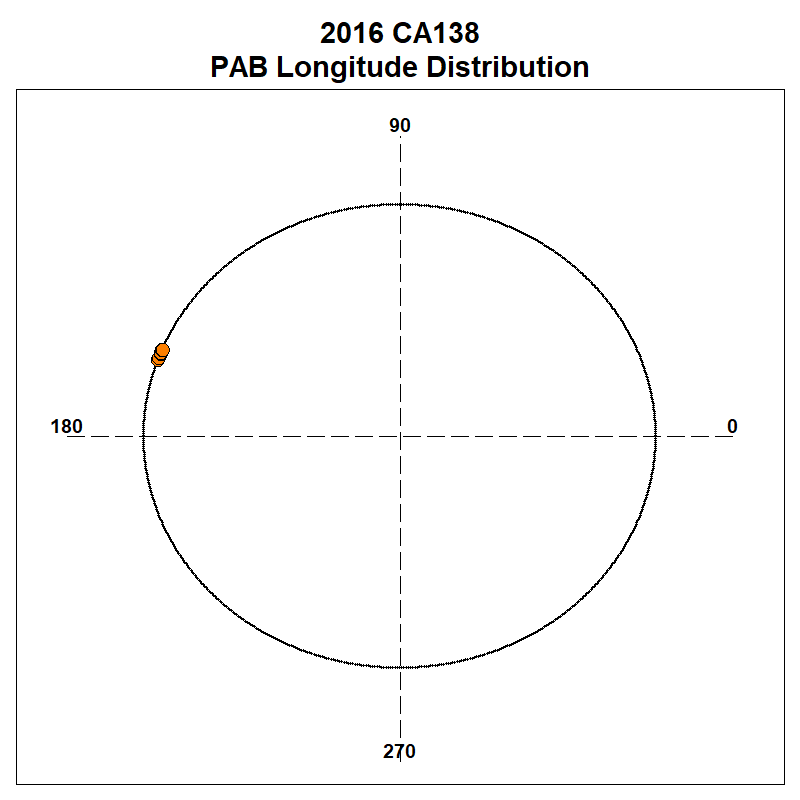}
    \includegraphics[width=0.5\columnwidth]{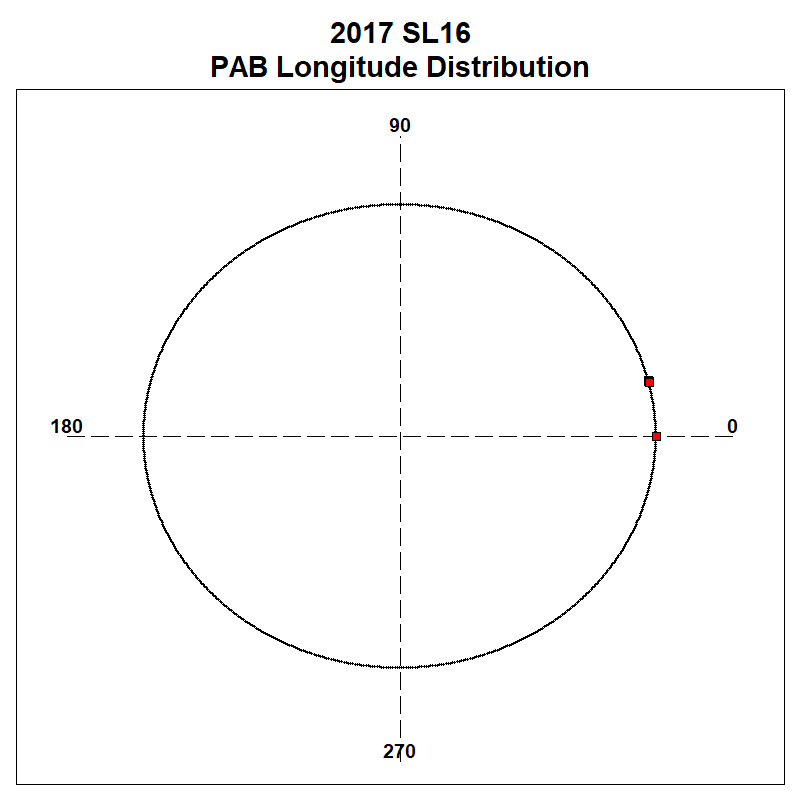}
	\caption{Phase angle distribution of all the observations (dense and sparse) for each of our four asteroids  (first row) and their corresponding phase angle bisector longitude (PABL) (second row)}% \doublespacing}
	\label{FIG:AllPhase}
	%\vspace{-0.5cm}
\end{figure*}
In addition, we are considering available sparse data on the asteroids  taken from the AstDys-2 database\footnote{\url{https://newton.spacedys.com/astdys/}}, choosing to use only those measurements with a  reported accuracy of 0.01 magnitude or higher (see Table~\ref{TBL:Sparse} for details). This helps us to extend the observational circumstances and especially the phase angle to a much larger range (see Figure~\ref{FIG:AllPhase} for details). Because sparse data is sporadic and with lower photometric accuracy than the dense data as they are mainly astrometry measurements and photometry is a secondary result, we are using those data with a lower weight of $0.3$ instead of the value of $1.0$ used for dense data \citep{Durech2009}. With such  a collection of dense and sparse data we can better determine the rotational period and the spin axis orientation. 
  
\section{Data reduction}
\subsection{Rotational period determination}
The data were reduced by applying standard bias and flat-field corrections followed by aperture photometry to produce the light curve for each of the objects.

To determine the rotational period of the asteroids, instead of resorting  to a standard Fourier analysis or investigating the $\chi^2$ of the fitted observational data by a Fourier function with different orders, we used a light curve inversion method to determine  a simple 3D shape model of the object that reproduces the modelled light curve. Then we compare this light curve to the observational data to obtain  the best period solution. For our purposes, we used the software provided by the Database of Asteroid Models from Inversion Techniques (DAMIT\footnote{\url{https://astro.troja.mff.cuni.cz/projects/damit/}}). 
 The software was developed by Mikko Kaasalainen in \texttt{\textbf{Fortran}} and converted to \texttt{\textbf{C}} by Josef Durech \citep{DAMIT}. For period determination we are using the sub-routine \texttt{\textbf{period\_scan}}.  To ensure that the global minimum of $\chi^2$ in the period search is not missed, we scan through a fairly wide interval  of possible periods.  Starting with six initial poles for each trial period and selecting the period that gives the lowest $\chi^2$, if there is a clear minimum in $\chi^2$ when plotted as a function of period we choose this solution as the best period.

If there is no clear minimum in the $\chi^2$ vs. period plot or that many pole solutions are giving the same residual – it means that there is not enough data for a unique model.

According to \citet{Kaasalainen2001}, the smallest separation $\Delta P$ of the local minima in the trial period $P$ spectrum of the $\chi^2$ of the light curve fit is roughly given by
\begin{eqnarray}\label{DP}
\frac{\Delta P}{P} \approx \frac{1}{2}\frac{P}{T}.
\end{eqnarray}
where $T={\rm max}(|t-t_0|)$ within the light curve set. i.e. the timespan of the observations.

\citet{Kaasalainen2001} also explain that the period uncertainty is mostly governed by the epochs of the light curves. If the best local $\chi^2$ minimum of the period spectrum is clearly lower than the others, one can obtain an error estimate of, say, a hundredth part of the smallest minimum width $\Delta P$ since the edge of a local minimum ravine always lies much higher than its bottom. Thus the period determination  can be very accurate for data that cover many years. On the other hand, if the neighbouring minima are not clearly higher than the best one, the accuracy cannot be considered better than $\Delta P$ since the local error estimate cannot be applied globally. 

\subsection{Spin axis orientation}\label{SEC:model}

 To determine a rotational pole solution for the asteroids, we ran the \texttt{\textbf{convexinv}} routine with different initial poles randomly distributed over the unit sphere and with 15 deg steps in both ecliptic longitude ($\lambda$) and latitude ($\beta$) to produce 312 initial pole orientations.  
 In such a way we are constructing a $\chi^2$-map using all the solutions (top panel of each of Fig.~\ref{FIG:CA138-pole},\ref{FIG:SL16-pole},\ref{FIG:WM64-pole}~\&~\ref{FIG:EE104-pole}) which we are using to isolate the areas with low $\chi^2$ and to find the best one. 
If there is one pole solution that gives significantly lower $\chi^2$ than all others, we adopt this pole solution. For asteroids orbiting near the ecliptic plane, there are always two possible poles with the same $\beta$ and $\lambda\pm$180. 

\section{Results}
\subsection{Objects with no previous period solution}
Here we are presenting results for objects in our sample  with no previously published rotational information.
\subsubsection*{\rm \bf 2017 SL16}
\begin{figure}[b!]
	\centering
	\includegraphics[angle=90,width=\columnwidth]{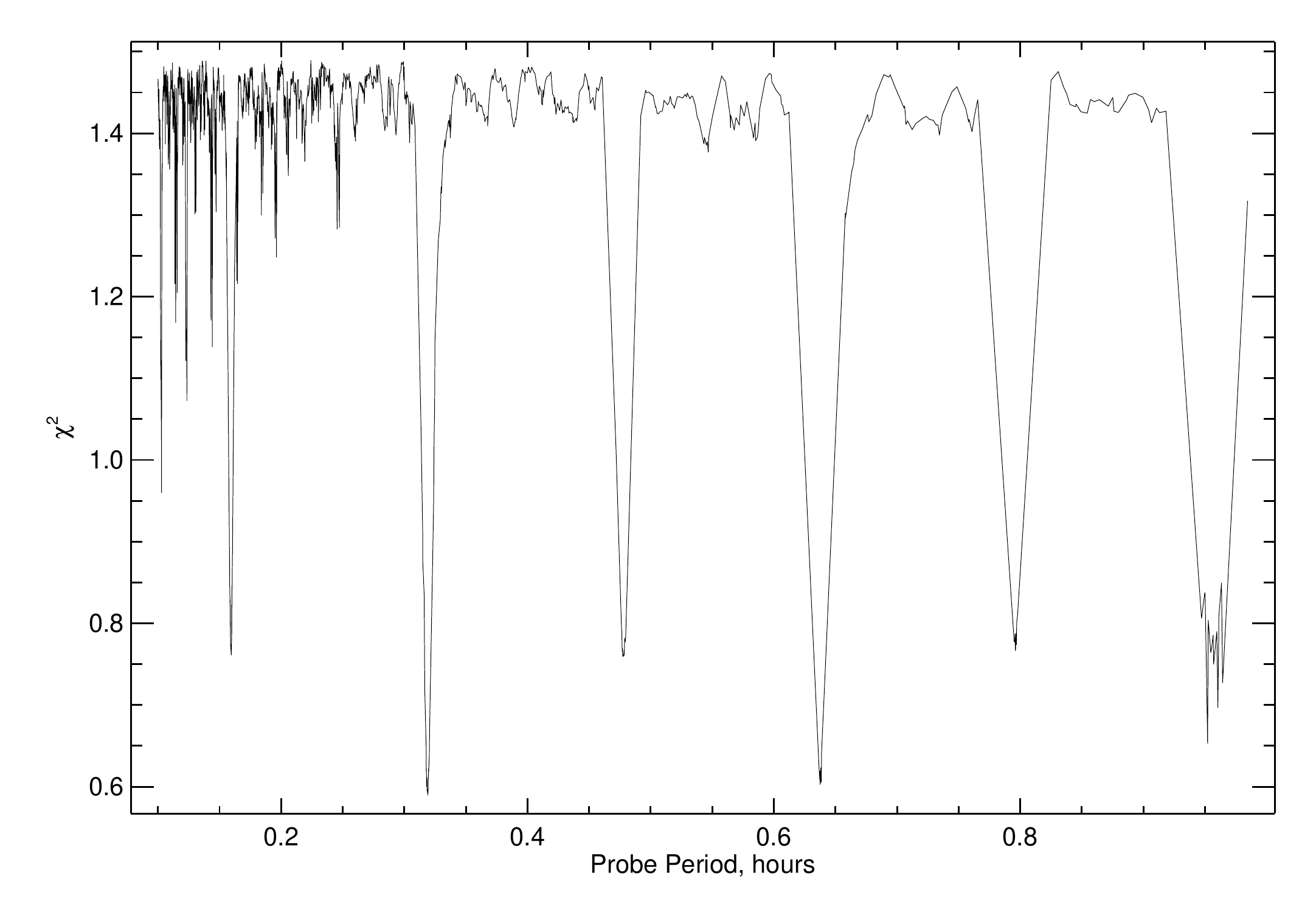}
    \includegraphics[angle=90,width=\columnwidth]{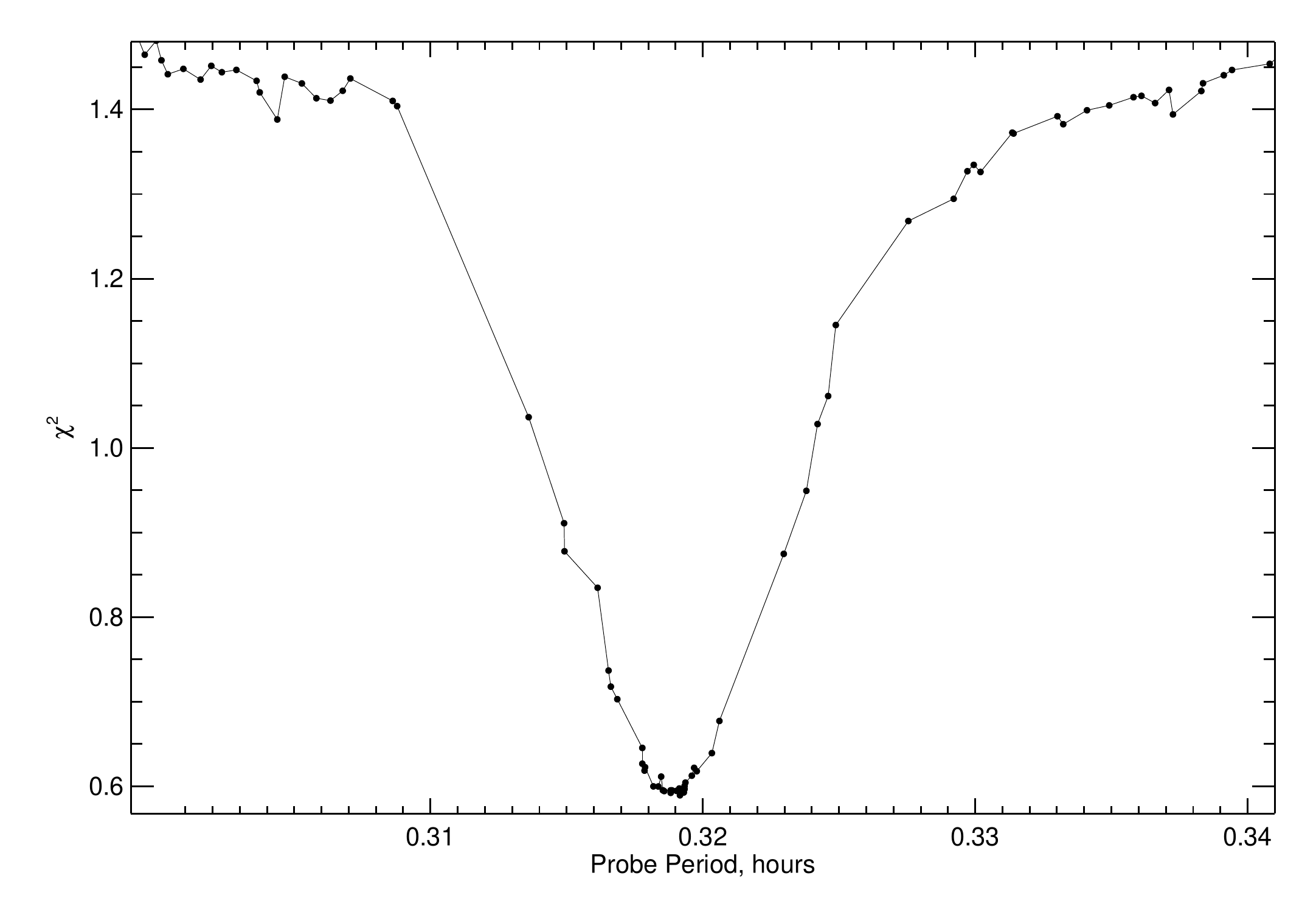}
	\caption{2017 SL16 - $\chi^2$ vs probe period plot used for a period search (top panel) and the same plot but zoomed around the best solution with the lowest $\chi^2$ value (bottom panel)}% \doublespacing}
	\label{FIG:SL16-per}
\end{figure}
\begin{figure}[t!]
	\centering
	\includegraphics[angle=90,width=\columnwidth]{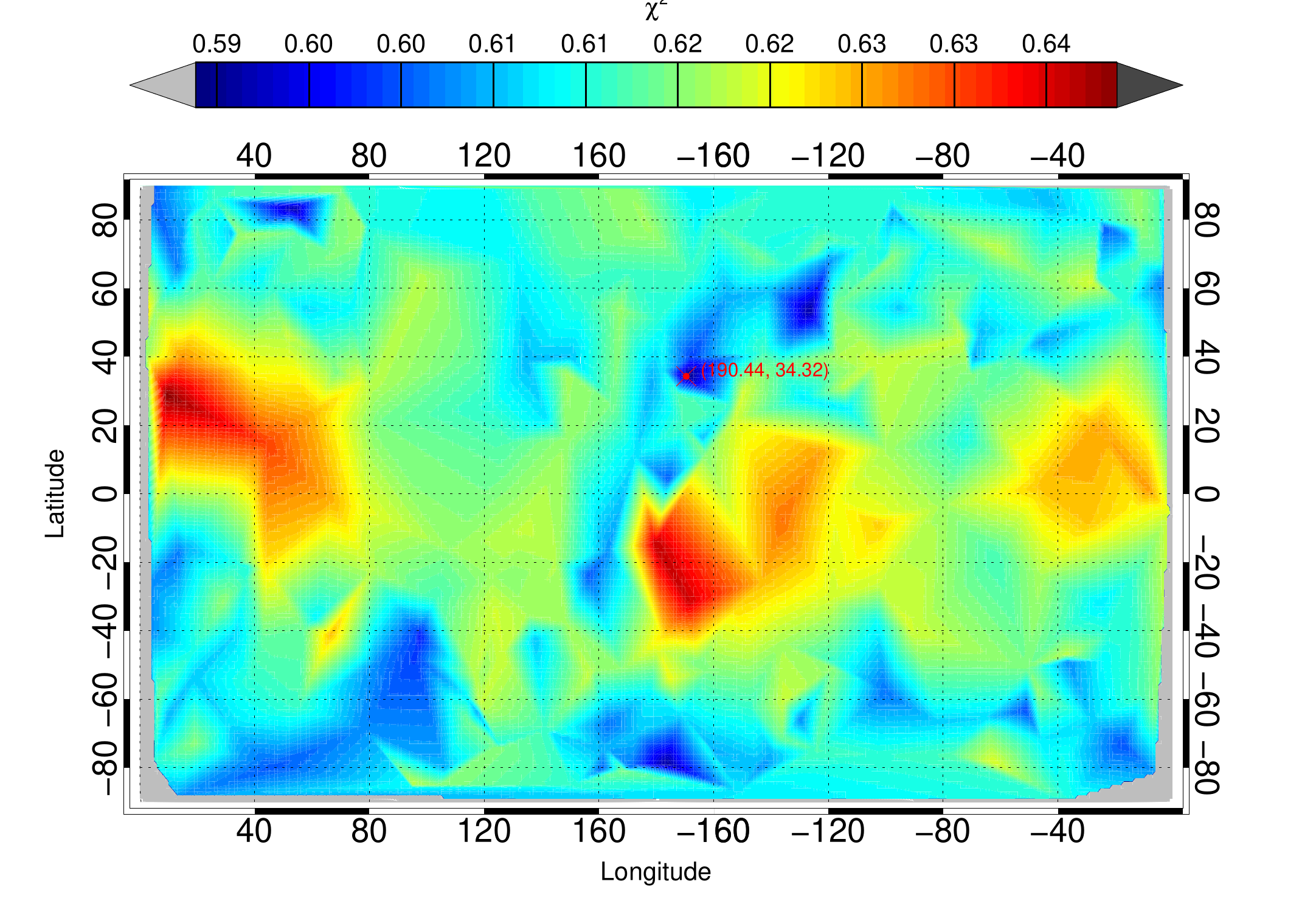}
    \includegraphics[angle=90,width=\columnwidth]{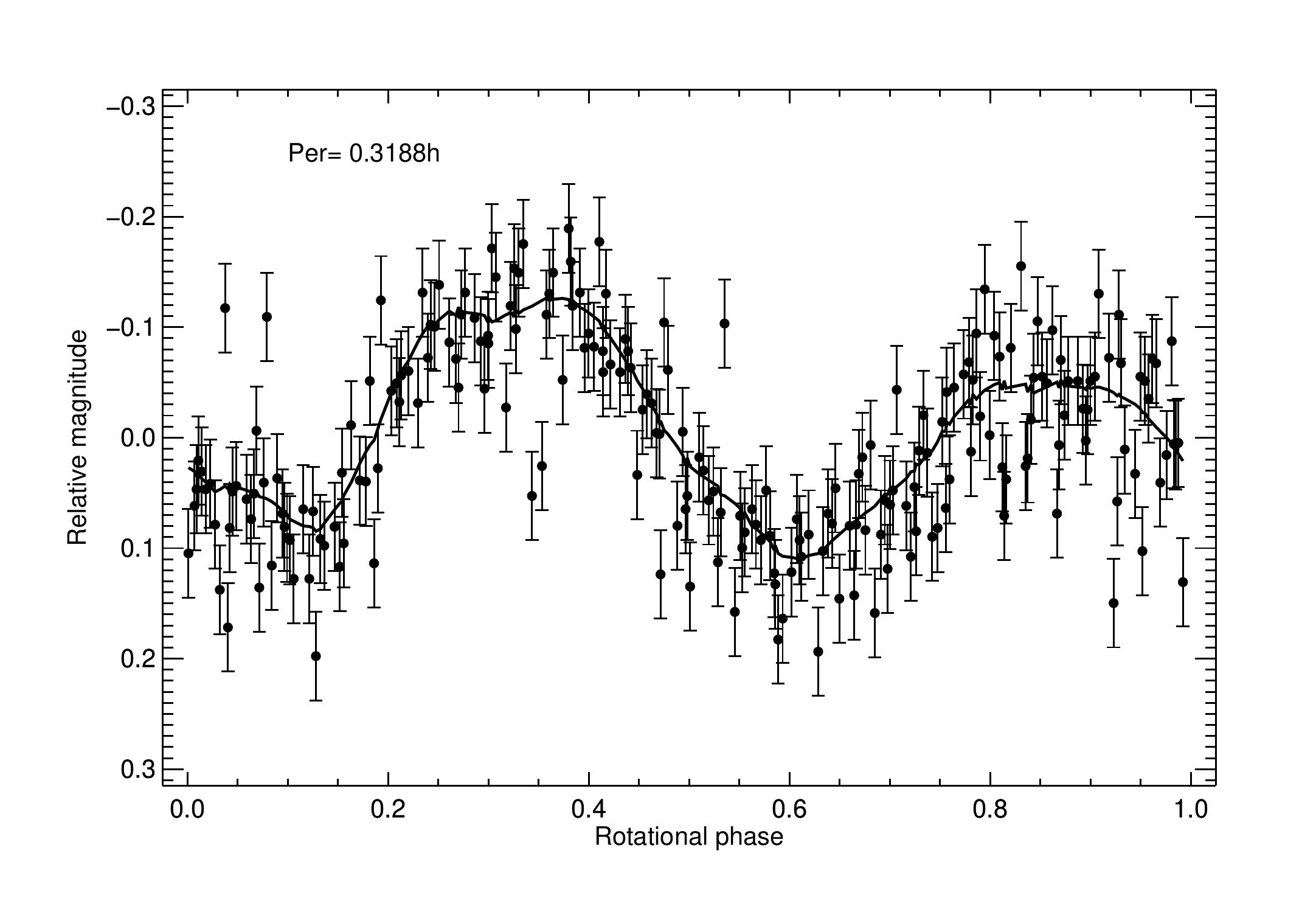}
    \includegraphics[angle=90,width=\columnwidth]{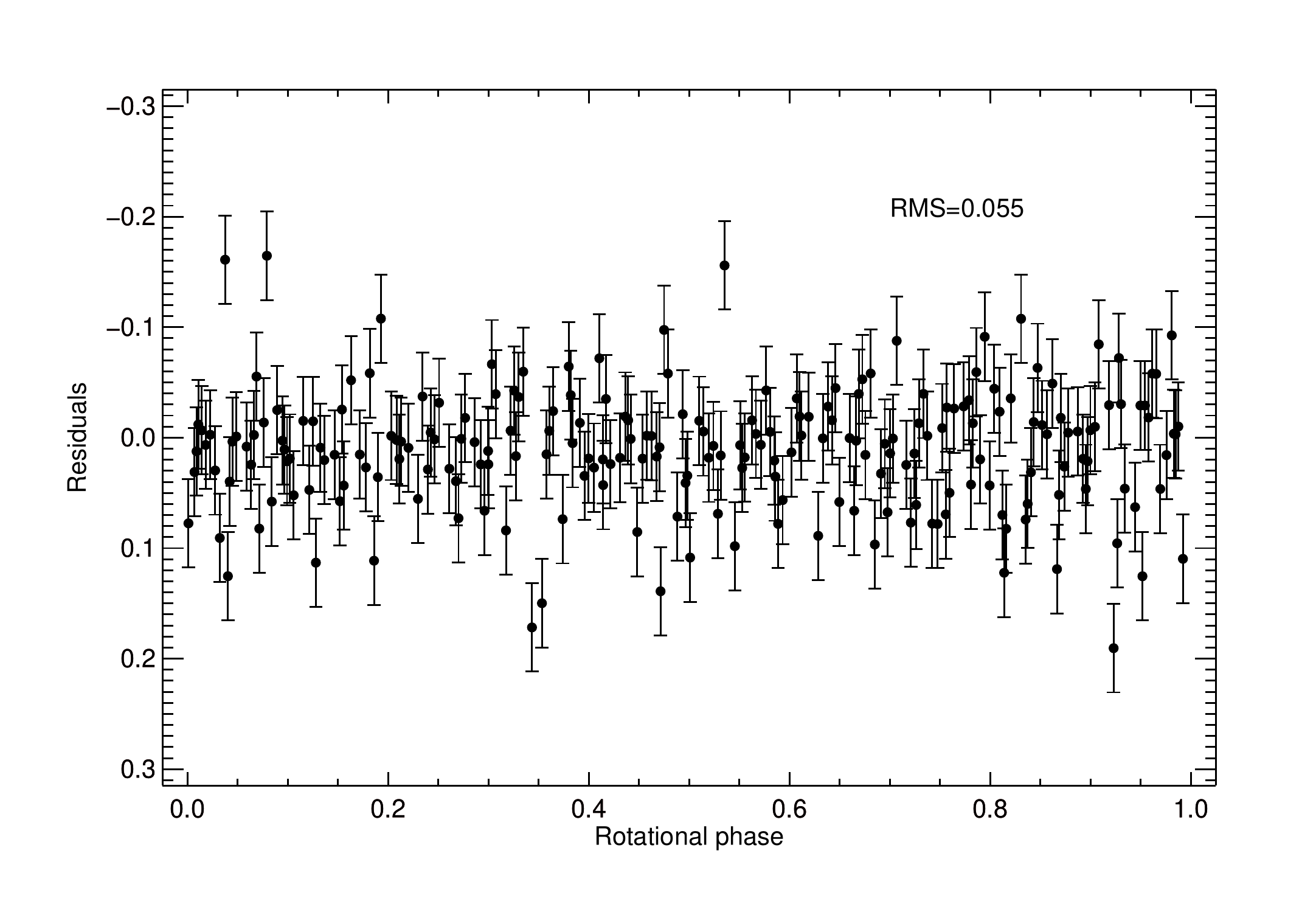}
	\caption{2017 SL16 - $\chi^2$-map of all 312 solutions described in section~\ref{SEC:model} and the solution with the lowest $\chi^2$ value marked with a red cross (top panel). The middle and bottom panels show the corresponding model fitting to the observational data and residuals, respectively. 
 }% \doublespacing}
	\label{FIG:SL16-pole}
%\vspace{-0.8cm}
\end{figure}

2017 SL16 (H=25.8) is the smallest object we observed based on its absolute magnitude. The light curve analysis shows that it is a very fast rotator. The $\chi^2$ variation with probe periods calculated  by  comparing the observations to the modelled light curves obtained from the simple 3D shape model is presented in Figure~\ref{FIG:SL16-per}. The period with the lowest $\chi^2$ is P=0.3188\,hrs ($\Delta$P=0.0053\,hrs). The other $\chi^2$ minima are multiples of the selected one,  our decision to adopt this solution  is based on the phased light curve having a shape with two minima and two maxima (middle panel of Figure~\ref{FIG:SL16-pole}).
%\newpage
Using the obtained rotational period we continue further by using it as an input value for a pole orientation search procedure. The result is shown at the top panel of Figure~\ref{FIG:SL16-pole} as a $\chi^2$-map which was constructed as described in section~\ref{SEC:model}. The corresponding model fitting to the observational data and residuals are presented on the middle and bottom panels of the same figure, respectively. The first five pole solutions  with $\chi^2$ less than 0.6  are presented in Table~\ref{TBL:Res}. We cannot make a conclusive decision which one is correct, but they all  have similar $\chi^2$ values and give very similar rotation periods and those marked "3" and "3m" could be mirror solutions as explained above.  Additional constraints on the rotational pole may we obtained from the $\chi^2$ map, in this case the pole solution is more likely to be far from the ecliptic than close to it. 
The $RMS$ for the entire dataset and for the model light curve with the lowest $\chi^2$ is $0.055$. 
%\vspace{-0.8cm}
\subsubsection*{\rm \bf 2016 CA138}
2016 CA138 (H=23.3) is an Aten asteroid and one of the faintest objects in our sample, with a diameter of D=75\,m assuming a visual geometric albedo p$_{\rm V}$=0.15. 
\begin{figure}[b!]
	\centering
		\includegraphics[angle=90,width=\columnwidth]{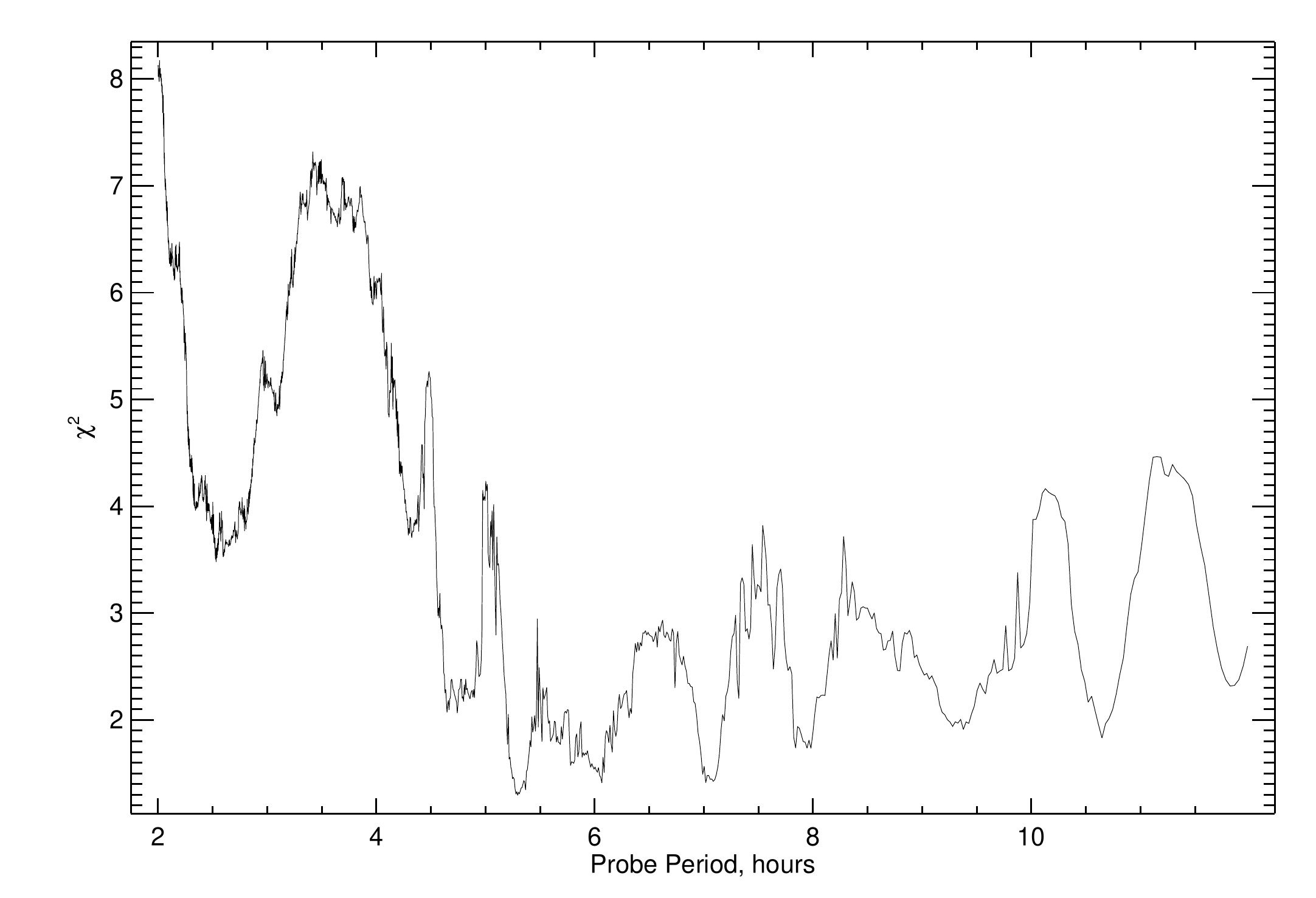}
\includegraphics[angle=90,width=\columnwidth]{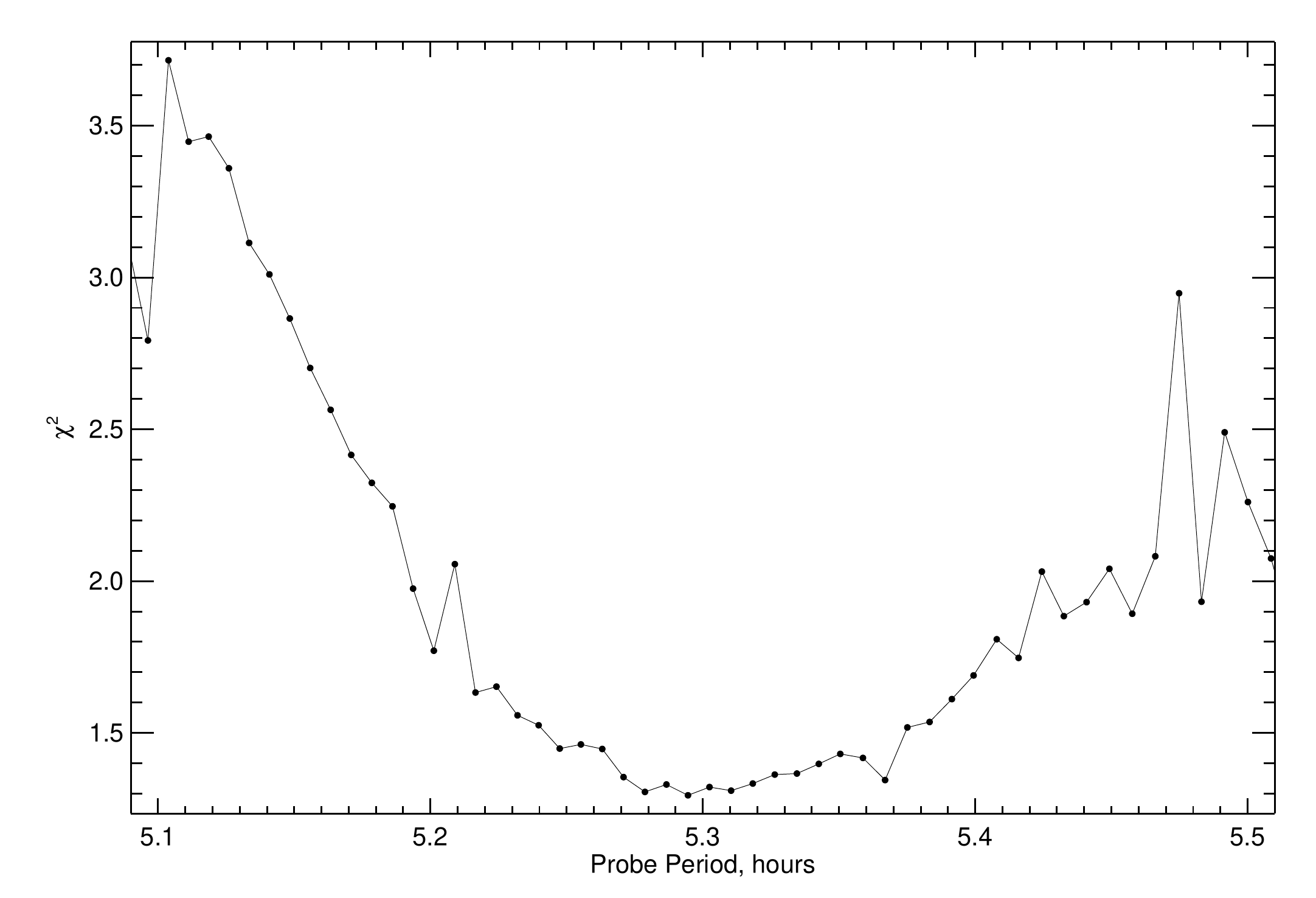}
	\caption{2016 CA138 - $\chi^2$ vs probe period plot used for a period search (top panel) and the same plot but zoomed around the solution with the lowest $\chi^2$ value (bottom panel)}% \doublespacing}
	\label{FIG:CA138-per}
\end{figure}

\begin{figure}[t!]
	\centering
		\includegraphics[angle=90,width=\columnwidth]{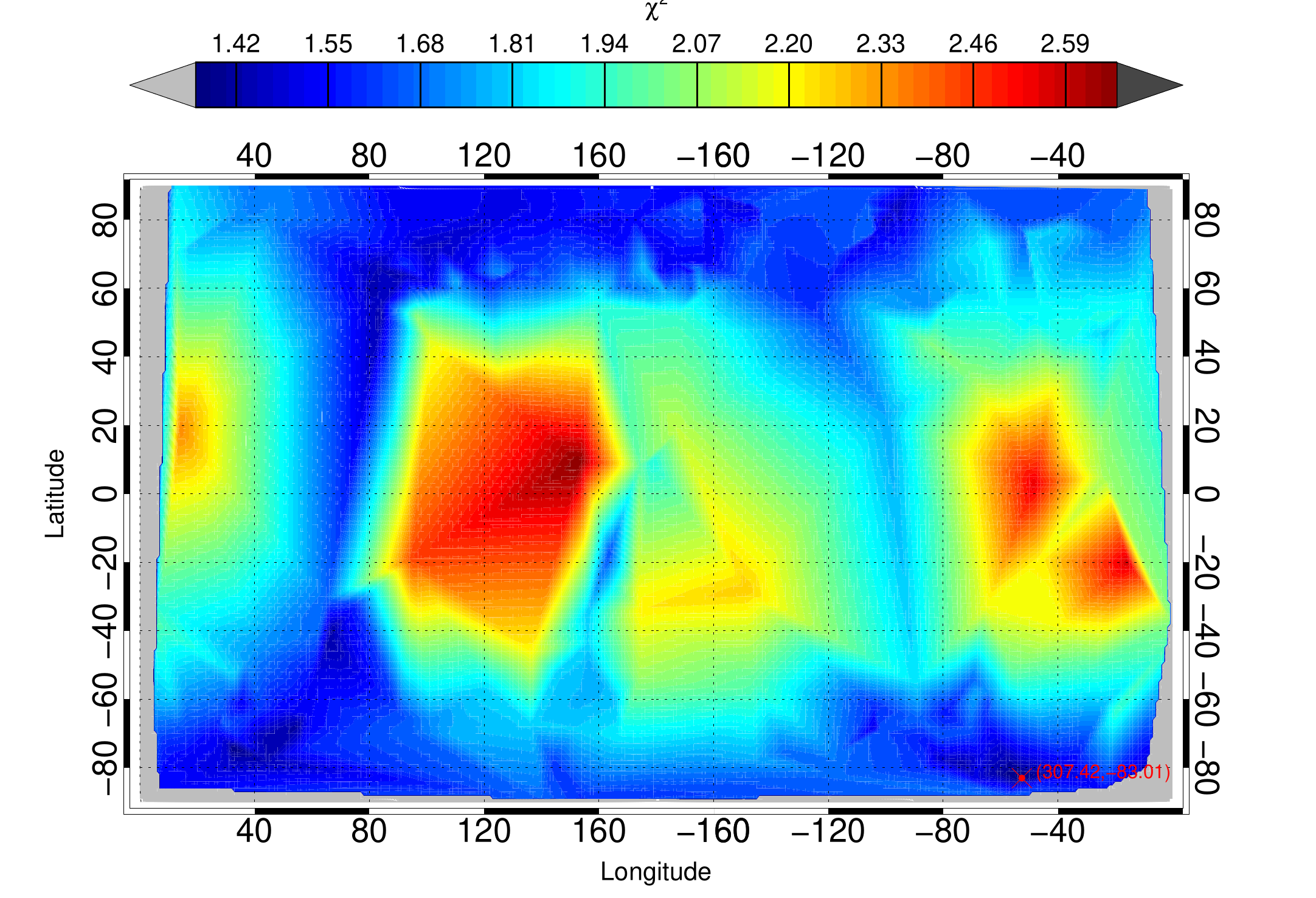}
\includegraphics[angle=90,width=\columnwidth]{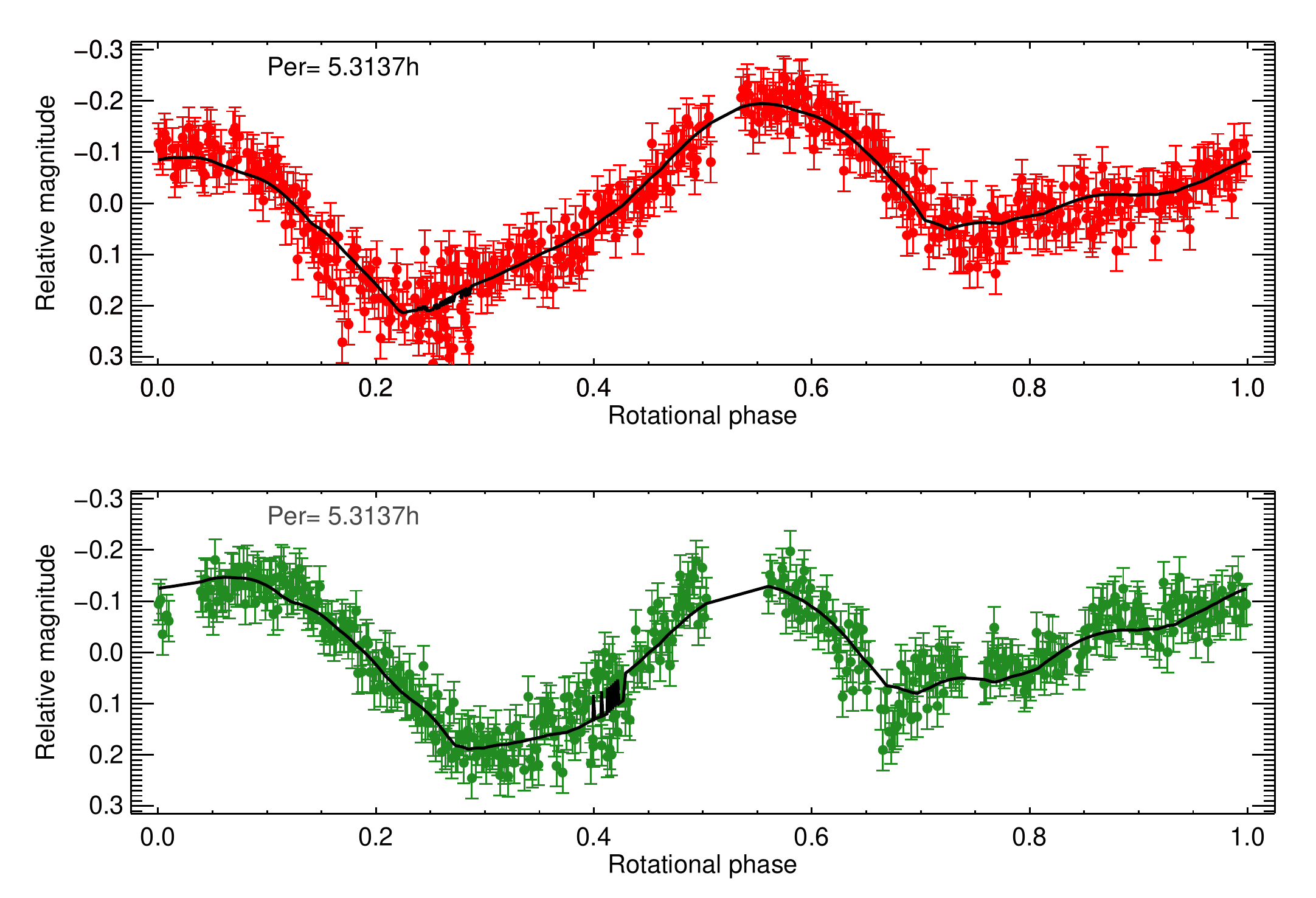}
\includegraphics[angle=90,width=\columnwidth]{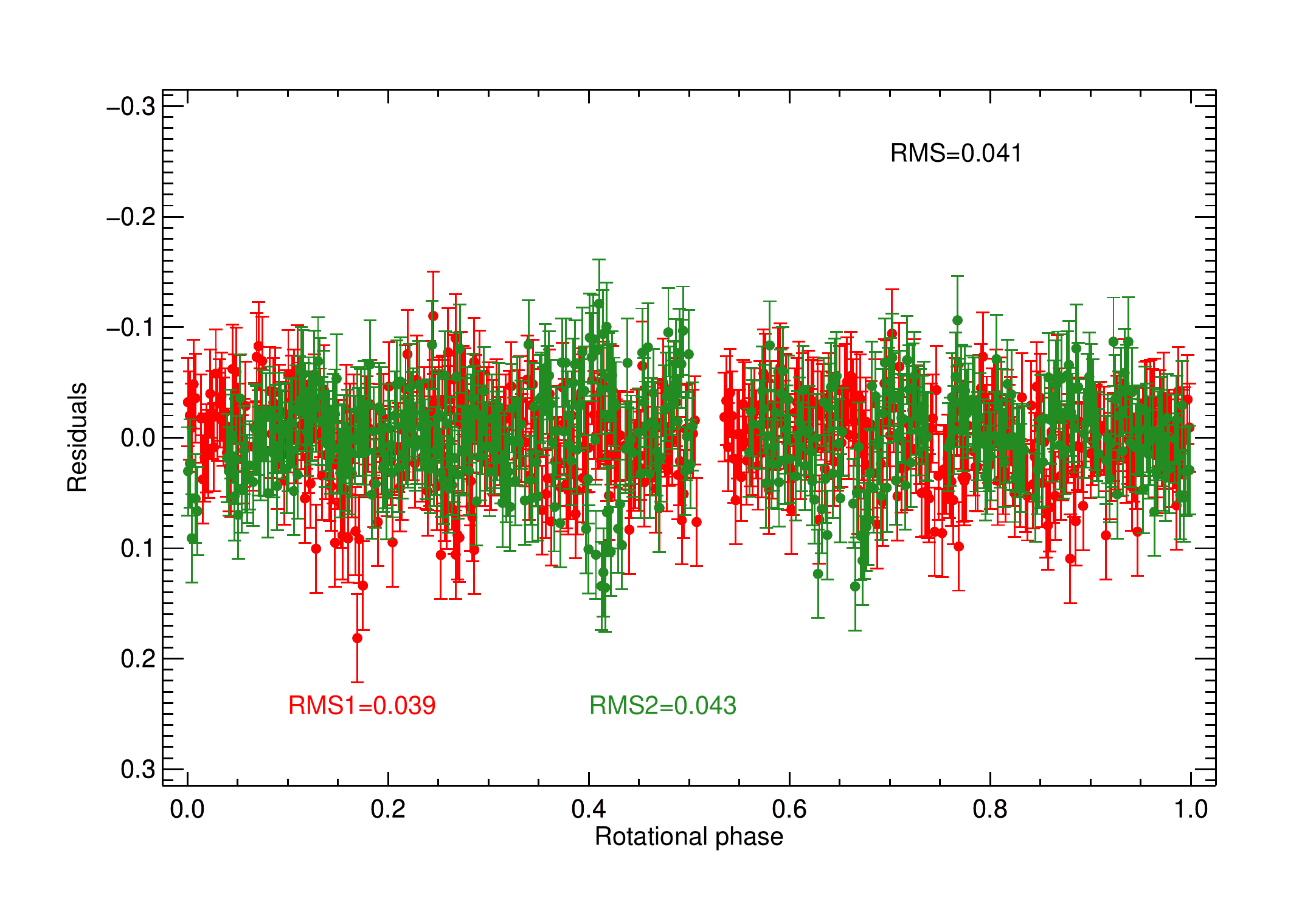}
	\caption{2016 CA138 - $\chi^2$-map of all 312 solutions (top panel) described in section~\ref{SEC:model} and the best solution with the lowest $\chi^2$ value marked with a red cross. The middle and bottom panels show the corresponding model fitted to the observational data and the fit residuals, respectively. Red and green colours represent the two different dates of observations - 17 and 18 February 2020, respectively.}% \doublespacing}
	\label{FIG:CA138-pole}
%\vspace{-0.8cm}
\end{figure}
No taxonomic or colour information is available for this object. Our solution for the rotational period (see Figure~\ref{FIG:CA138-per}) is P=5.3137\,hrs ($\Delta$P=0.0016 hrs). 

The $\chi^2$-map of the pole solutions is presented in the top panel of Figure~\ref{FIG:CA138-pole}. It indicates that the asteroid rotational pole should be close to the ecliptic pole but with two exceptions - the vertical trenches along 60 and 260 degrees longitude.  The solution with the deepest $\chi^2$ minimum is at pole coordinates \mbox{$\lambda_1$=307.42$^{\circ}$} and \mbox{$\beta_1$=-83.01$^{\circ}$}.
However, any other solution inside the trenches - which are almost 180 degrees apart in longitude and could be interpreted as mirror solutions - are plausible.  
The $RMS$ for the entire dataset and the model light curve with the lowest $\chi^2$ is $0.043$. 
%\vspace{-0.8cm}
\subsection{Objects with previous period solutions}
Here we are presenting {improved rotational state solutions for objects with previous spin rates published in \citet{Borisov2021}.}

\subsubsection*{\rm \bf (418849) 2008 WM64}
(418849) 2008 WM64 (H=20.6) is an Apollo asteroid. It has a previously published rotation period of P=2.40$\pm$0.02\,h \citep{Rowe2018,Warner2018a}. 
 We combined our sample of dense and sparse data together with dense data from \citeauthor{Rowe2018} consisting of 21 light curves in 'R' band. We note that these latter data do not improve the PABL distribution of the observations as they were performed over a single night 4 days earlier than ours. We then carry out a fit as for the asteroids using this combined dataset. 
\begin{figure}[b!]
%\vspace{-0.8cm}
	\centering
		\includegraphics[angle=90,width=\columnwidth]{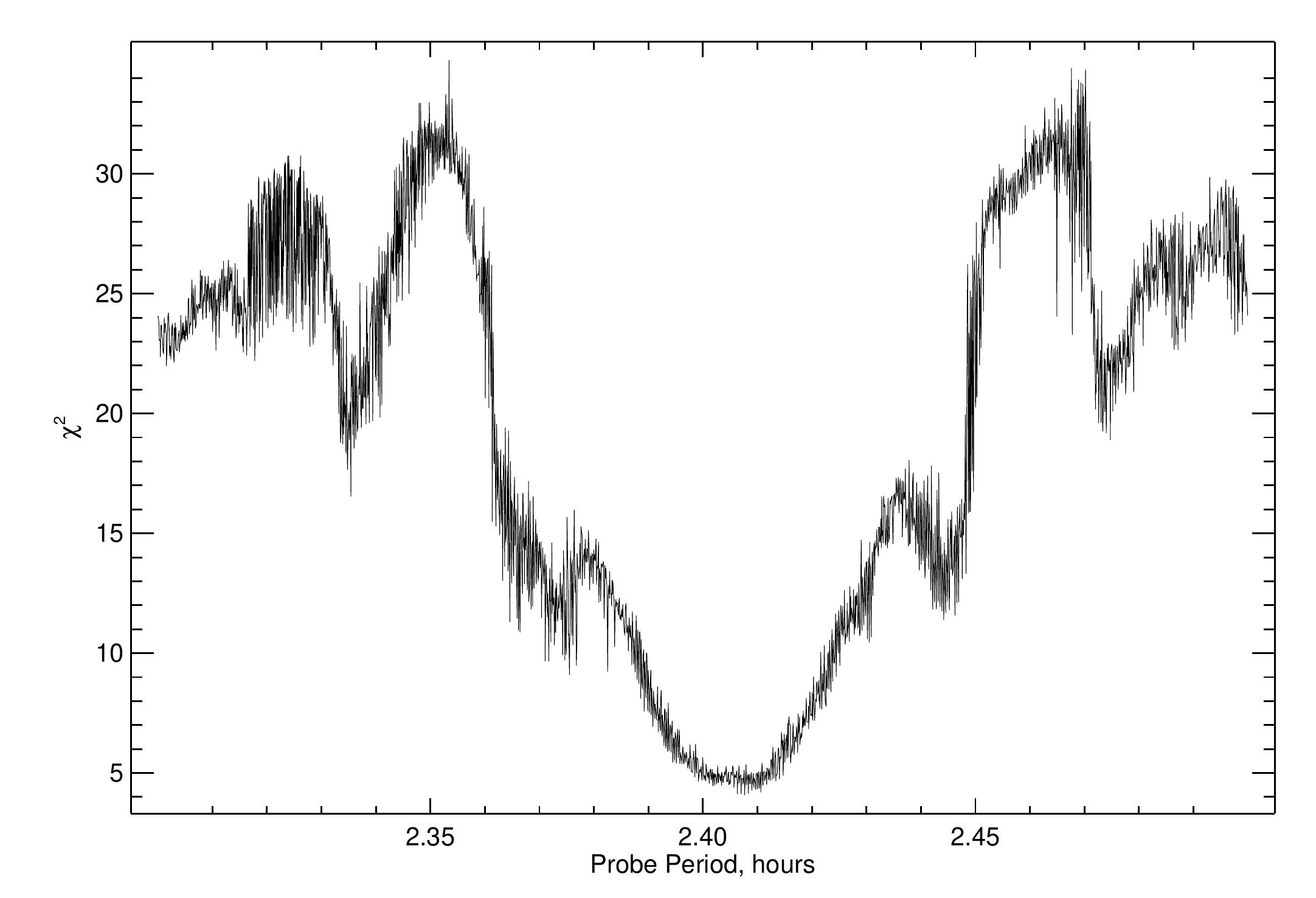}
\includegraphics[angle=90,width=\columnwidth]{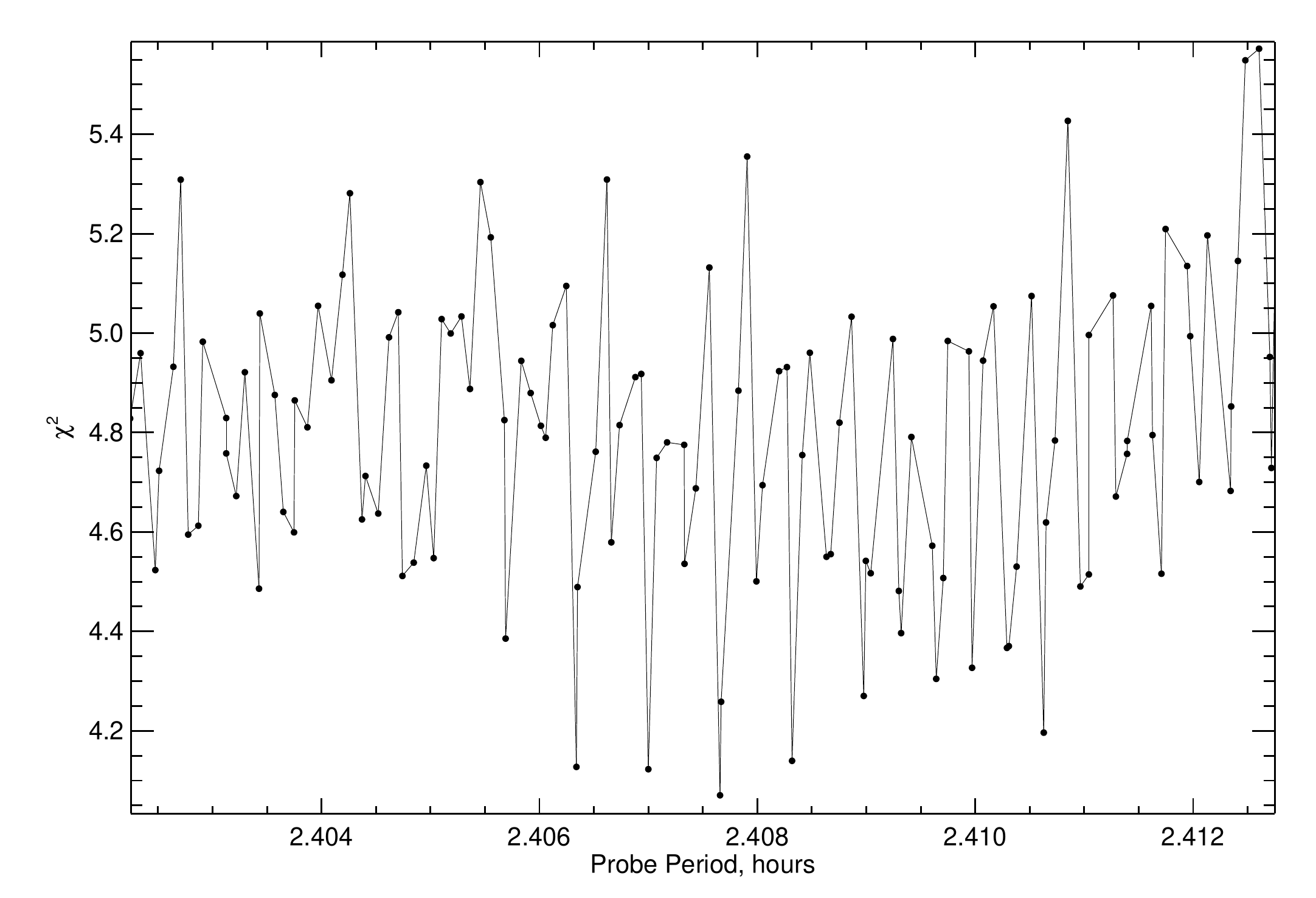}
	\caption{(418849) 2008 WM64 - $\chi^2$ vs probe period plot used for a period search (top panel) and the same plot but zoomed around the best solution with the lowest $\chi^2$ value (bottom panel)}% \doublespacing}
	\label{FIG:WM64-per}
\end{figure}
%\newpage
\begin{figure}[t!]
	\centering
	\includegraphics[angle=90,width=\columnwidth]{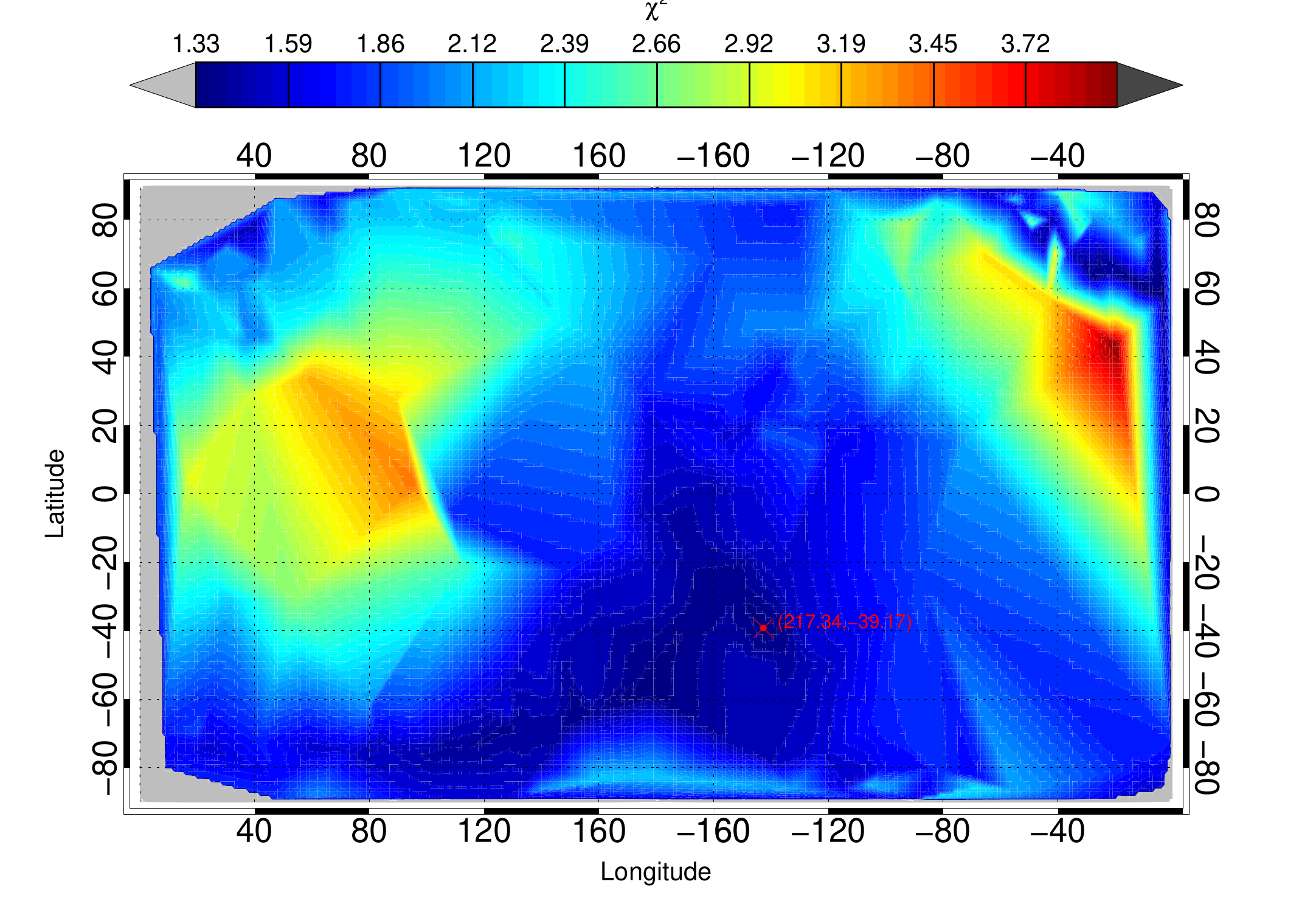}
    \includegraphics[angle=90,width=\columnwidth]{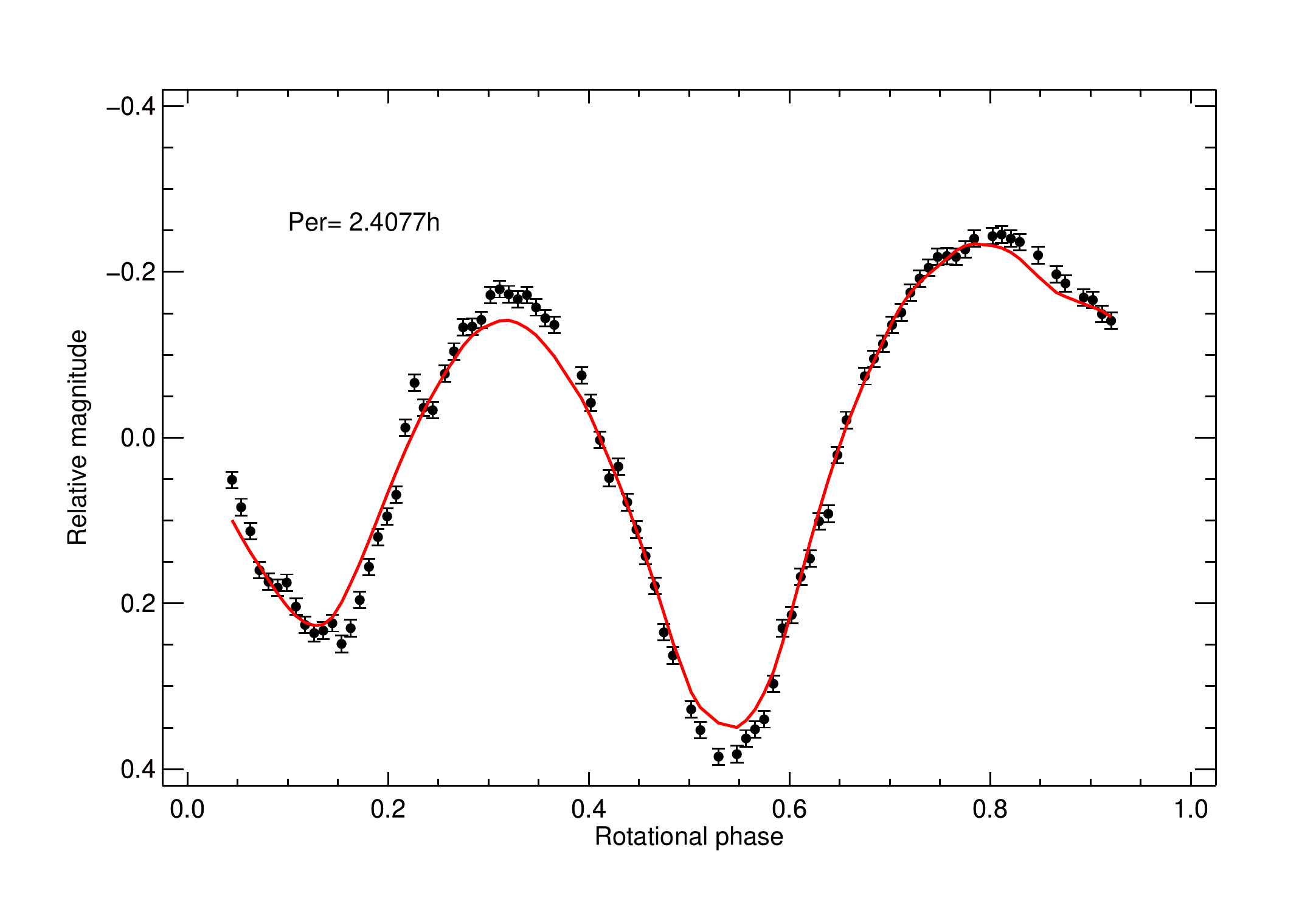}
    \includegraphics[angle=90,width=\columnwidth]{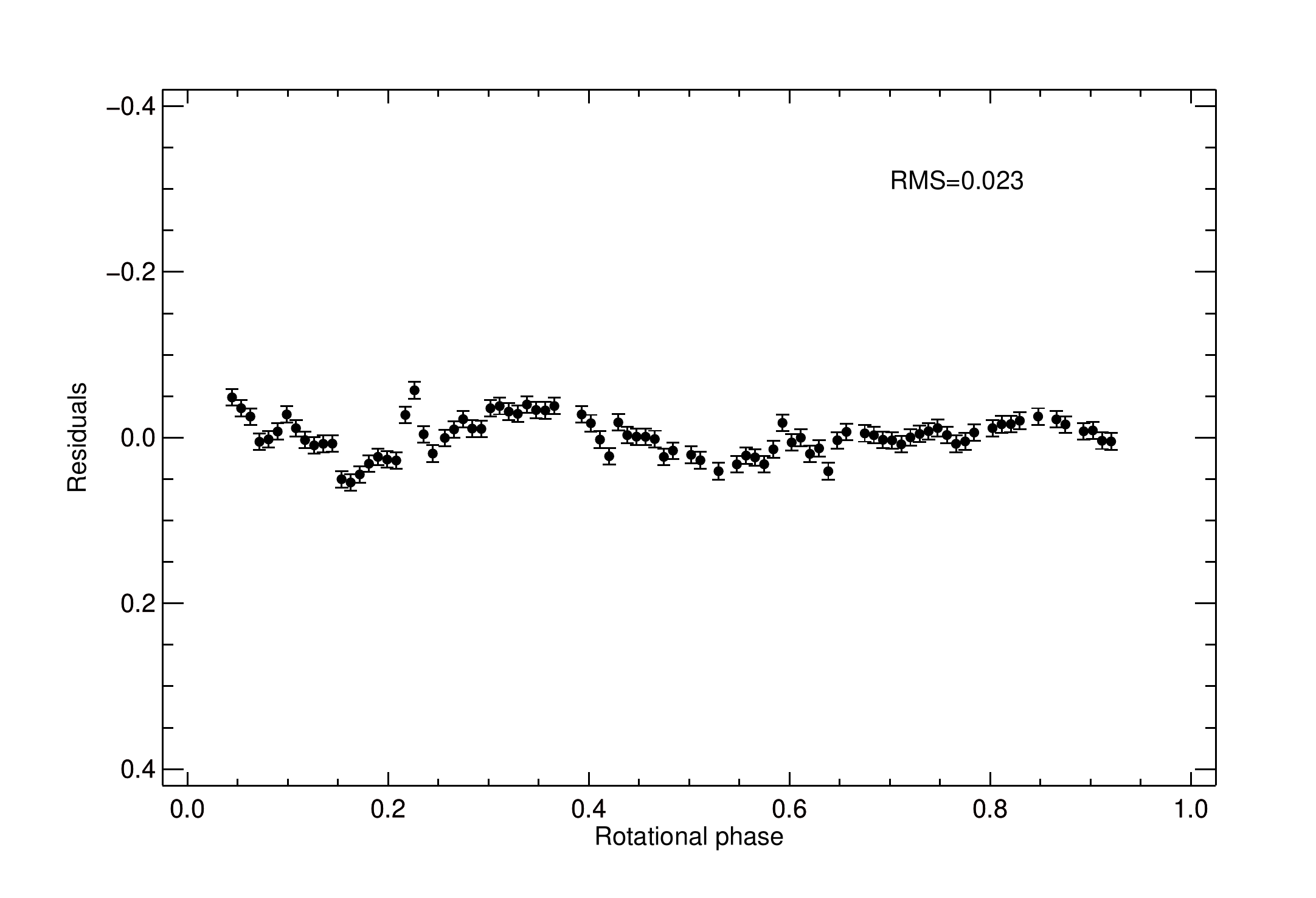}
	\caption{(418849) 2008 WM64 - $\chi^2$-map of all 312 solutions described in section~\ref{SEC:model} and the best solution with the lowest $\chi^2$ value marked with a red cross (top panel). The middle and the bottom panels show the corresponding model fit to the observational data and residuals, respectively.}% \doublespacing}
	\label{FIG:WM64-pole}
\end{figure}

Our rotational period search, presented in Figure~\ref{FIG:WM64-per}, yields an asteroid spin rate of P=2.4077\,hrs ($\Delta$P=0.0001 hrs), slightly larger than in our previous work \citep[][P=2.356$\pm$0.033\,hrs]{Borisov2021} but closer to the solution obtained by \citet{Rowe2018}. 
The $\chi^2$-map in Fig.~\ref{FIG:WM64-pole} indicates a pole direction south of the ecliptic with formal best estimate at \mbox{$\lambda_1$=217.34$^{\circ}$} and \mbox{$\beta_1$=-39.17$^{\circ}$}.  The $RMS$ for the entire dataset and the model light curve with the lowest $\chi^2$ is $0.045$. 

\subsubsection*{\rm \bf (138175) 2000 EE104}
\begin{figure}[b!]
	\centering
		\includegraphics[angle=90,width=\columnwidth]{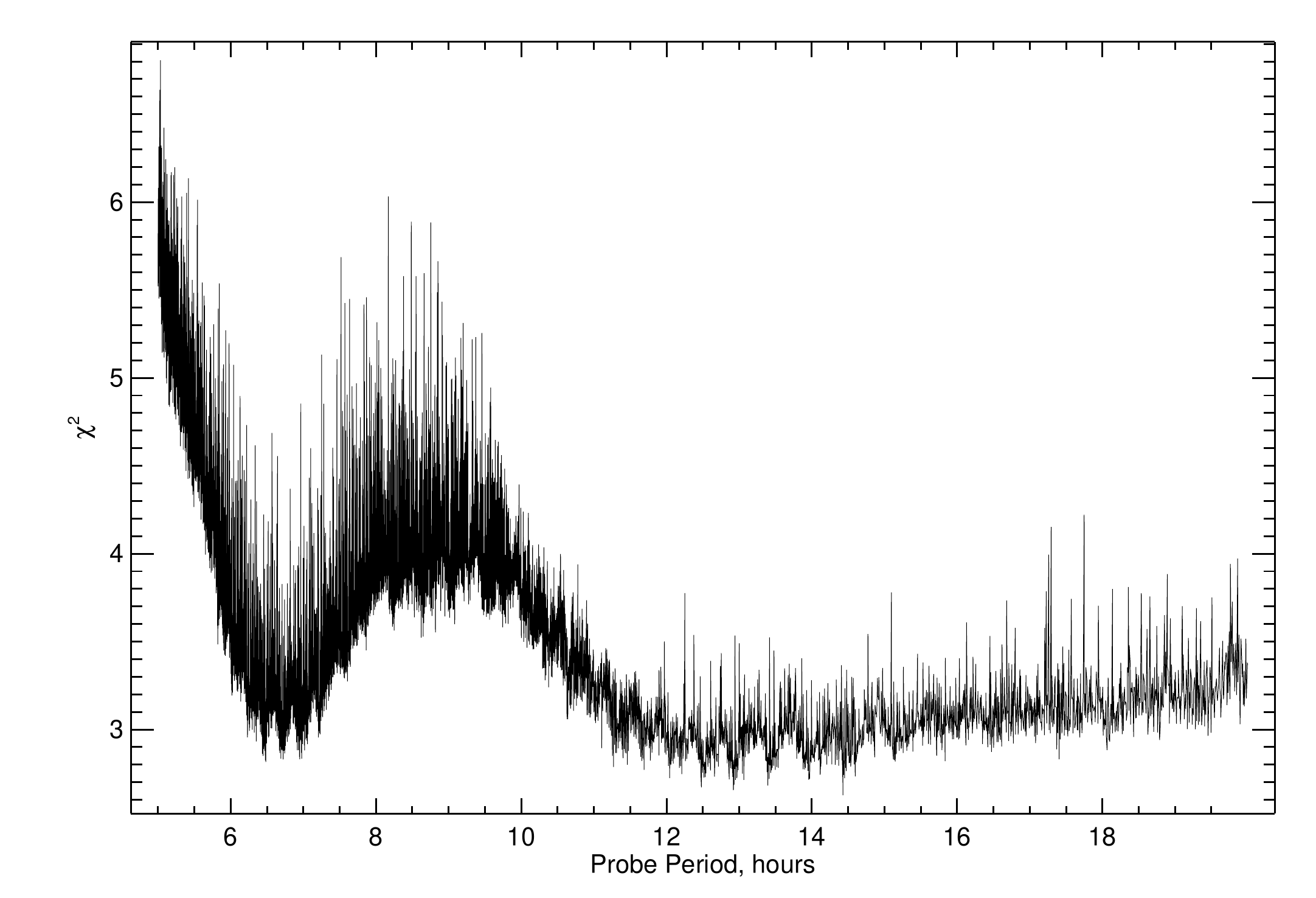}
\includegraphics[angle=90,width=\columnwidth]{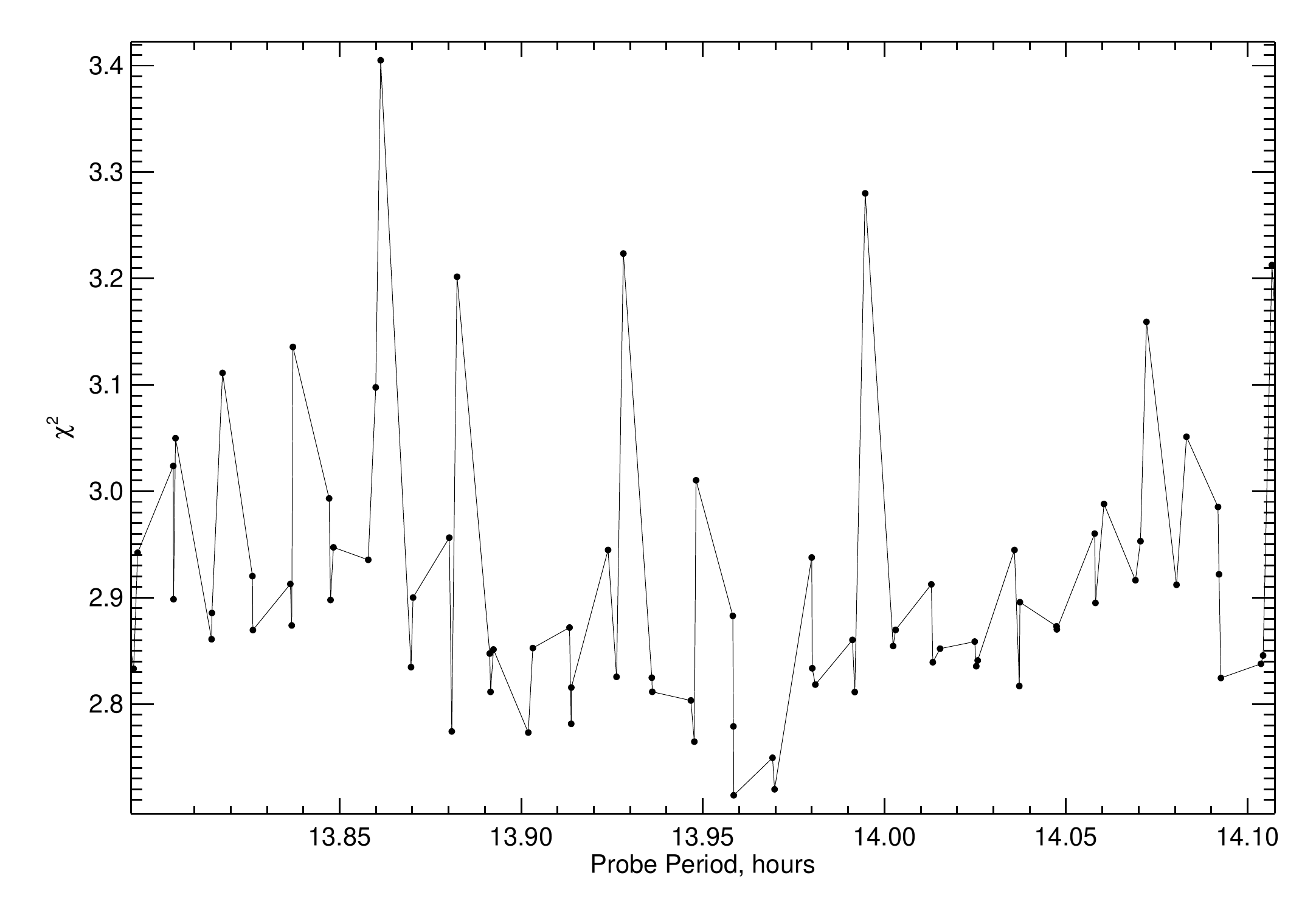}
	\caption{(138175) 2000 EE104 - $\chi^2$ vs probe period plot used for a period search (top panel) and the same plot but zoomed around the solution with the lowest $\chi^2$ value (bottom panel)}% \doublespacing}
	\label{FIG:EE104-per}
\end{figure}

(138175) 2000 EE104 (H=20.4) is an Apollo asteroid  suspected to have debris spread along its orbit due to detection of interplanetary magnetic field disturbances near the orbital nodes \citep{Lai2017}. \citeauthor{Lai2017} argue that this material can be boulders with diameters larger than 10\,m. \citep{Jewitt2020} finds no evidence for co-moving companions or a dust particle trail and report a B-V colour of 1.16$\pm$0.04 which they interpret as intermediate between C-class and S-class asteroids. The mean B-R colour is consistent with that measured for Jovian Trojan and D-type asteroids.

Asteroids can shed material from their surface and onto heliocentric orbit if they rotate once every few hours or faster \citep{Pravec2010,Jacobson2011}. 
\begin{figure}[t!]
	\centering
		\includegraphics[angle=90,width=\columnwidth]{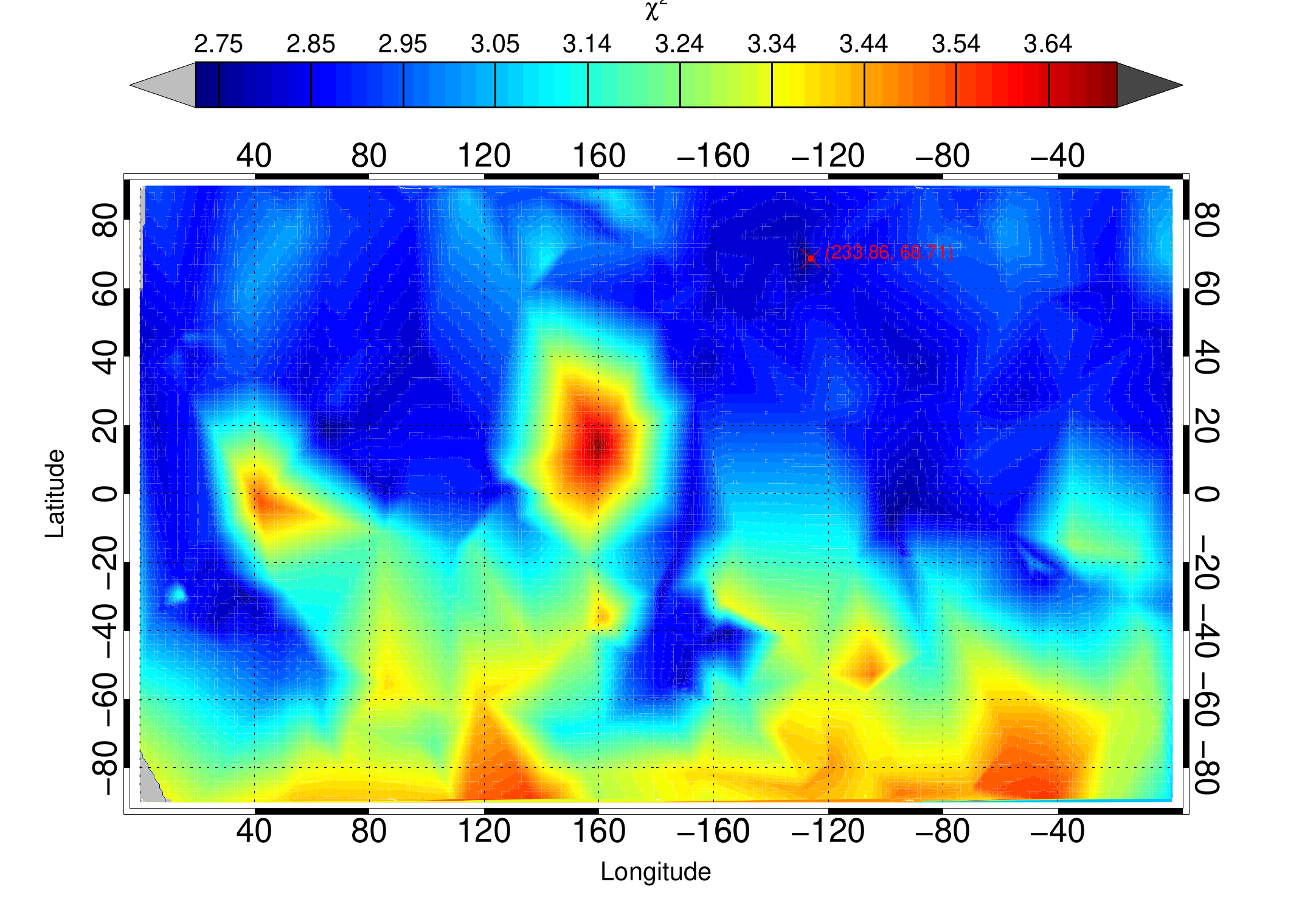}
\includegraphics[angle=90,width=\columnwidth]{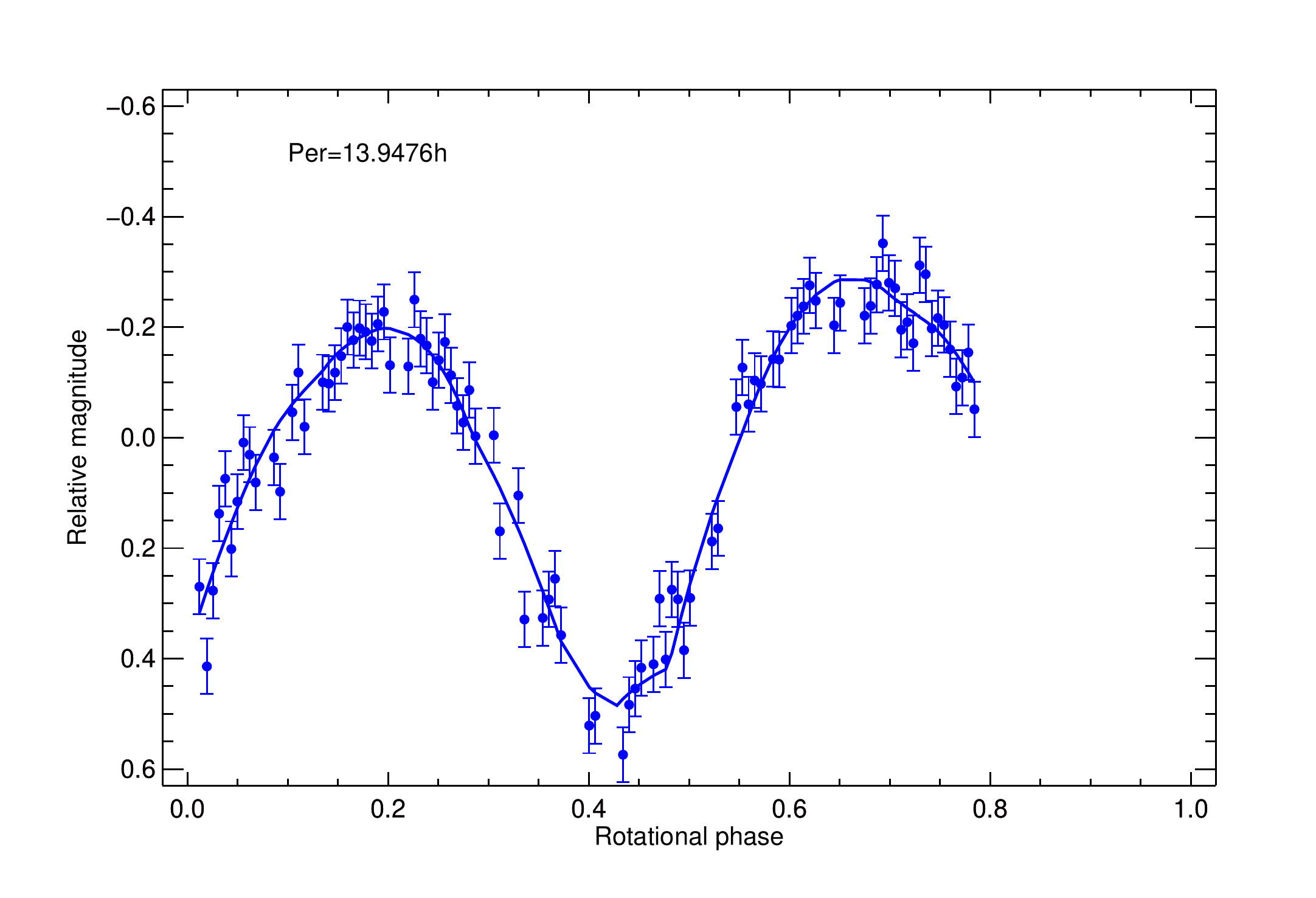}
\includegraphics[angle=90,width=\columnwidth]{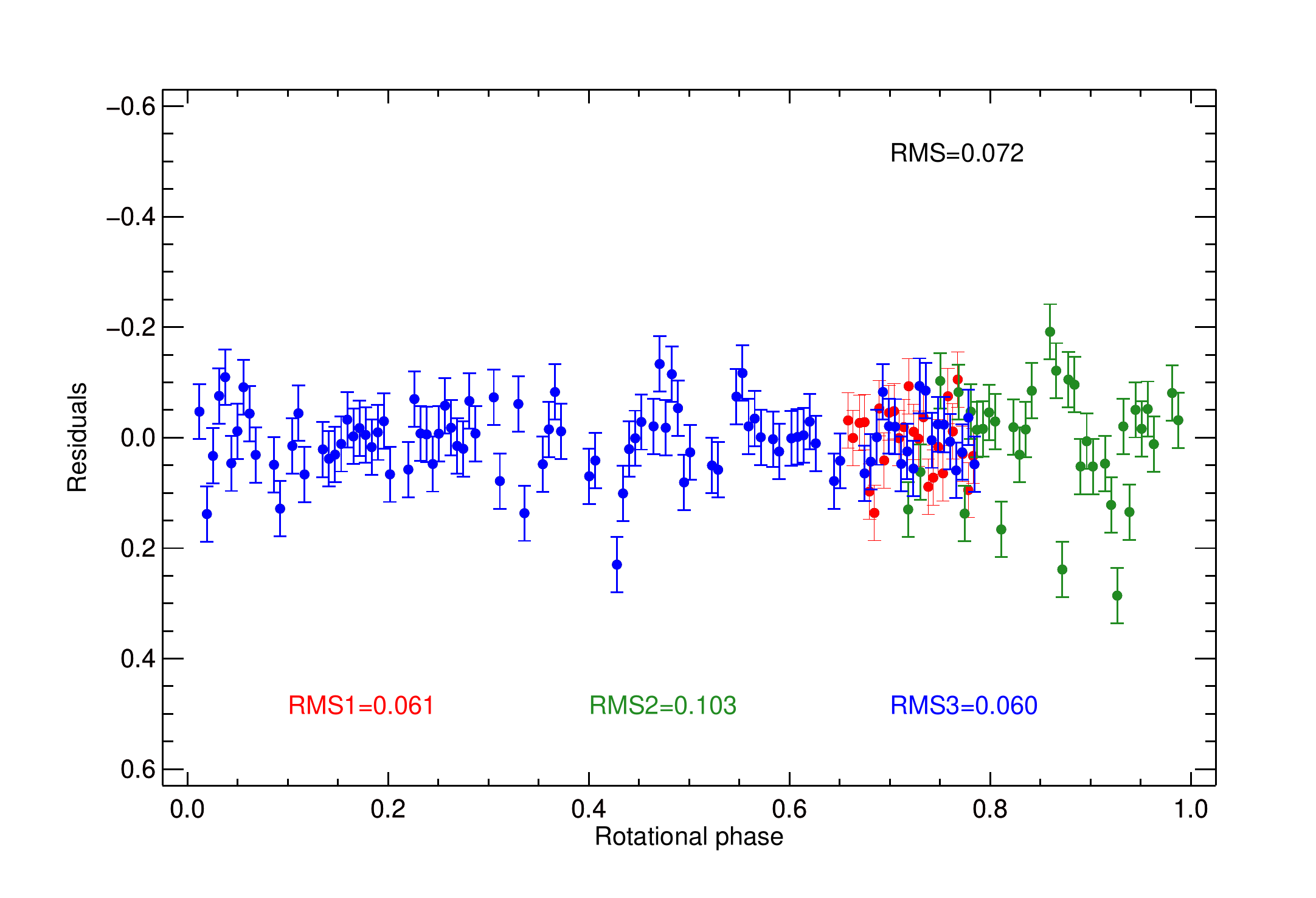}
	\caption{(138175) 2000 EE104 - $\chi^2$-map of all 312 solutions described in section~\ref{SEC:model} and the best solution with the lowest $\chi^2$ value marked with a red cross (top panel). The middle and the bottom panels show the corresponding model fitting to the observational data and residuals, respectively. {Red, green and blue colours represent the three different dates of observations - 09 November 2018, 01 January 2019 and 02 January 2020, respectively.}}% \doublespacing}
	\label{FIG:EE104-pole}
\end{figure}
Our investigation shows that this asteroid is a relatively slow rotator with a period of about 14\,hrs. Our rotational period and pole solution estimates are presented in Figure~\ref{FIG:EE104-per} and Table~\ref{TBL:Res}. A notable feature in this Figure is a lower $\chi^2$ for pole solutions north of the ecliptic and a single, broad minimum  at pole coordinates \mbox{$\lambda_1$=233.86$^{\circ}$} and \mbox{$\beta_1$=68.71$^{\circ}$} with corresponding period \mbox{P$_1$=13.9476\,hrs}. However, we cannot dismiss the two mirror solutions with similar $\chi^2$ values south of the ecliptic, also well defined as $\chi^2$ minima in Figure~\ref{FIG:EE104-pole} and presented in Table~\ref{TBL:Res}.
The $RMS$ for the entire dataset and the model light curve with the lowest $\chi^2$ is $0.127$. 

\subsection{Comparison to NEAs}
\begin{figure}[b!]
\flushright
\includegraphics[width=\columnwidth]{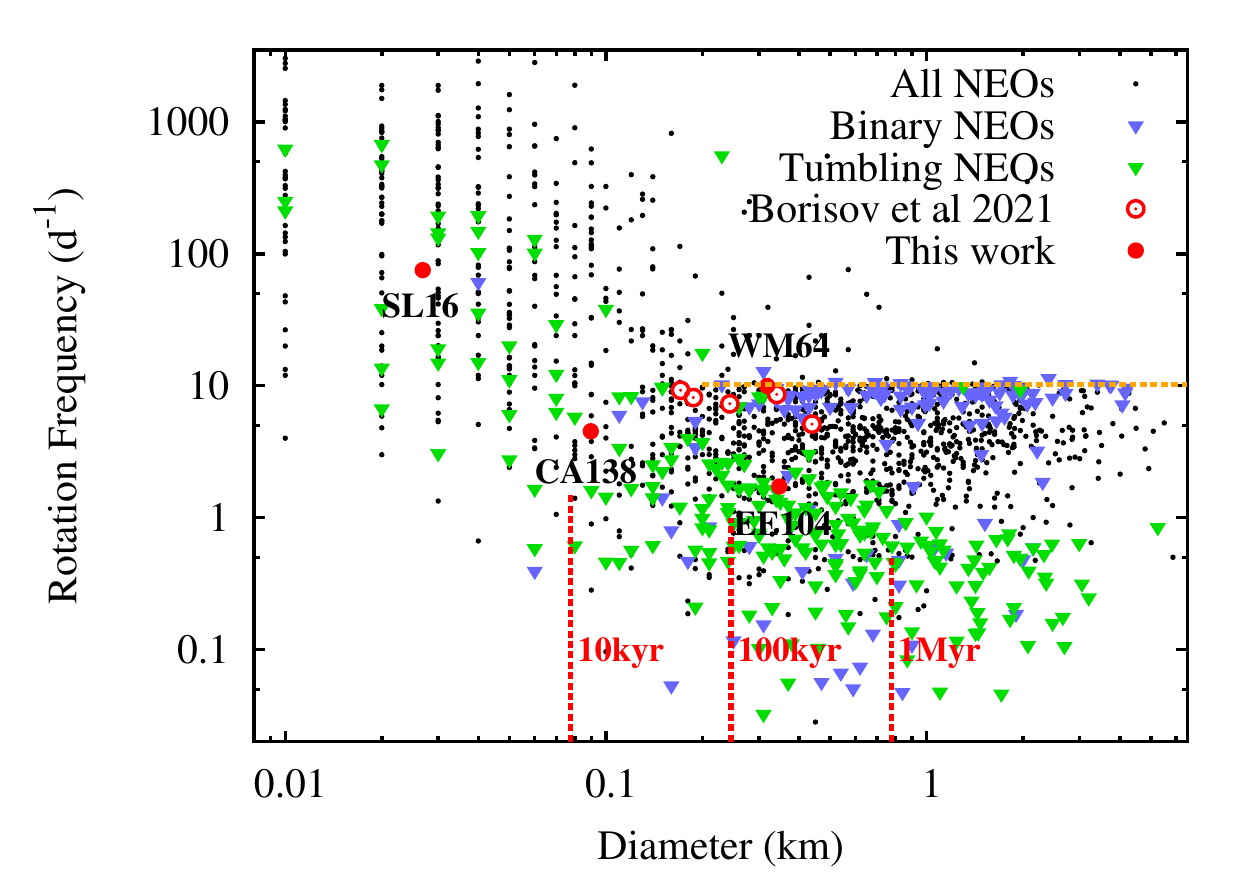}
\caption{Size vs spin rate of co-orbital asteroids in our sample (red points) compared to NEAs entries in the Asteroid Lightcurve Data Base (LCDB Bundle v4.0) as of December 2021, including confirmed binary and tumbling asteroids (blue and green points resp.). The horizontal line corresponds to the critical spin rate $\omega_{crit}(\rho)$ for $\rho =$ 2000\,kg\,m$^{-3}$. Different YORP spinup times for the asteroids are shown as vertical red lines. }% \doublespacing} 
\label{FIG:DiaPer2}
\end{figure}

 In Figure~\ref{FIG:DiaPer2} we compare the bulk properties of the four asteroids in our primary sample (filled circles) to NEAs entries in the LCDB database \cite[][black points]{lcdb_v4_2021} including confirmed binary systems (blue symbols) and objects in non-principal-axis or ``tumbling'' rotational state (green symbols). Our inferred sizes of sample asteroids are averages of two estimates assuming typical albedo values for C- and S-type asteroids 
\cite[$p_{C}=0.057\pm0.013$ and $p_{S}=0.197\pm0.051$;][]{Pravec.et.al2012}.  The vertical lines in the figure indicate YORP-induced spinup timescales $\tau_{Y}$ of $0.01$, $0.1$ and $1$\,Myr resp.~using the expression $\tau_{Y}\simeq1.68 D^{2}$\,Myr  from \citet{Jacobson2014} appropriate for NEAs, where $D$ is the diameter in km. For a prograde rotator, we expect the YORP torque to evolve the asteroid towards a critically-spinning configuration in <1\,Myr. 

\begin{table*}[width=1.0\textwidth,cols=15,pos=t!]
\caption{Results from searching for period  and  pole solutions  for all four objects.}\label{TBL:Res}
\begin{minipage}{1.0\textwidth}\centering
\begin{tabular*}{\tblwidth}{@{} RLLCCCCCRRCCCCC@{} }
\toprule
\multicolumn{2}{c}{Asteroid} & Sol.\footnote{Solution number (m - if it is a mirror solution)} &  \multicolumn{1}{c}{Period} & \multicolumn{1}{c}{$\Delta$P} & \multicolumn{1}{c}{MnMx$_{\rm data}$\footnote{The difference between the minimum and the maximum of the data}} & \multicolumn{1}{c}{A$_{\rm data}$\footnote{The amplitude of the data computed as a difference between the five point average around the minimum and the maximum of the data}} & \multicolumn{1}{c}{A$_{\rm model}$\footnote{The amplitude of the modeled light curve}} & \multicolumn{2}{c}{Pole coo} & \multicolumn{3}{c}{Axis ratio} & \multicolumn{1}{c}{Dark\footnote{The dark facet area in \%}} & \multicolumn{1}{c}{$\chi^2$} \\
(Number) & \multicolumn{1}{r}{Designation} & \# & \multicolumn{1}{c}{(hrs)} & \multicolumn{1}{c}{(hrs)} & \multicolumn{1}{c}{(mag)} & \multicolumn{1}{c}{(mag)} & \multicolumn{1}{c}{(mag)} & $\lambda(^{\circ})$ & $\beta(^{\circ})$ & a/b & a/c & b/c & \multicolumn{1}{c}{(\%)} &  \\
\midrule
(418849) & 2008 WM64 & 1 & 2.4077 & 0.0001 & 0.63 & 0.60 & 0.59 & 217.34 & -39.17 & 1.24 & 1.46 & 1.18 &  0.03 & 1.32529\\
\\
(138175) & 2000 EE104 & 1  & 13.9476 & 0.0051 & 1.01 & 0.84 & 0.77 & 233.86 &  68.71 & 1.20 & 2.98 & 2.49 & 0.01 & 2.72627 \\
(138175) & 2000 EE104 & 2  & 13.9471 & 0.0051 & 1.01 & 0.84 & 0.84 &  35.79 & -30.20 & 1.43 & 2.31 & 1.61 & 0.02 & 2.73148 \\
(138175) & 2000 EE104 & 2m & 13.9473 & 0.0051 & 1.01 & 0.84 & 0.78 & 205.24 & -40.27 & 1.63 & 3.44 & 2.10 & 0.02 & 2.74698 \\
\\
 & 2017 SL16 & 1  & 0.3188 & 0.0053 & 0.39 & 0.25 & 0.24 & 190.44 &  34.32 & 2.16 & 2.62 & 1.21 & 0.00 & 0.59162 \\
 & 2017 SL16 & 2  & 0.3190 & 0.0053 & 0.39 & 0.25 & 0.23 & 183.66 & -78.85 & 1.13 & 2.40 & 2.12 & 0.01 & 0.59326 \\
 & 2017 SL16 & 3  & 0.3192 & 0.0053 & 0.39 & 0.25 & 0.22 & 233.61 &  53.61 & 1.68 & 1.85 & 1.10 & 0.00 & 0.59345 \\
 & 2017 SL16 & 3m & 0.3192 & 0.0053 & 0.39 & 0.25 & 0.22 &  53.61 &  83.48 & 1.18 & 1.61 & 1.36 & 0.00 & 0.59345 \\
 & 2017 SL16 & 4  & 0.3191 & 0.0053 & 0.39 & 0.25 & 0.24 &  97.24 & -41.24 & 1.33 & 4.06 & 3.06 & 0.00 & 0.59948 \\
\\
& 2016 CA138 & 1  & 5.3137 & 0.0016 & 0.56 & 0.45 & 0.41 & 307.42 & -83.01 & 1.24 & 3.46 & 2.78 & 0.12 & 1.36563 \\
& 2016 CA138 & 2  & 5.3141 & 0.0016 & 0.56 & 0.45 & 0.40 &  32.23 & -73.89 & 1.04 & 2.56 & 2.46 & 0.01 & 1.38376 \\
& 2016 CA138 & 3  & 5.3139 & 0.0016 & 0.56 & 0.45 & 0.41 & 263.00 &  83.00 & 1.19 & 2.62 & 2.21 & 0.43 & 1.45168 \\
& 2016 CA138 & 3m & 5.3139 & 0.0016 & 0.56 & 0.45 & 0.41 &  83.00 &  66.45 & 1.11 & 2.68 & 2.42 & 0.43 & 1.45168 \\
\bottomrule
\end{tabular*}
\end{minipage}
\end{table*}

 In order to increase the sample size, we have included five Earth co-orbital asteroids from \citet{Borisov2021}: (468909) 2014 KZ44, (468910) 2014 KQ76, (512245) 2016 AU8, (522684) 2016 JP and 2018 EB (open circles) where available data allows to estimate the spin period but not produce a full rotational state. We then refer to these nine asteroids as the extended sample to distinguish from the original sample of four asteroids. 
 
 Overall, co-orbital asteroids in the extended sample appear to have rotation rates similar to NEAs of similar size. Several objects in our sample, including 2008 WM64, cluster near the  critical rotation frequency $\omega_{crit} \approx 2 \pi \left(3 \pi /\rho G\right)^{-1}$  where a strengthless ``rubble pile'' body of bulk density $\rho$ would begin to come apart \cite[see, e.g.,][and references therein]{HarrisPravec2005}, a regime populated by the smallest primaries within the binary asteroid population.
 
Arguably, the most noteworthy case is that of 2017 SL16, with an estimated $D=20-36$\,m the smallest asteroid in our sample. Its fast, $\sim$15 minute rotation period suggests an asteroid held together by internal strength. The next smallest object is 2016 CA138 ($D=66-121$\,m) with a spin period of $\sim$5\,hr, rather slow but still within the observed range for objects of similar size. A more extreme, though by no means exceptional, case is the $\sim$14\,hr period of 2000 EE104, ($D=255-470$\,m), slower than most asteroids in the same size range. This asteroid, along with 2016 CA138 and 2017 SL16, is located in the regime occupied by tumbling asteroids although we note here that no evidence of a secondary period or the presence of satellites was found in our photometric data for these asteroids.

\section{Numerical Simulations}
\subsection{Simulation setup}
To investigate the orbit evolution of the asteroids, we used the {\sc hybrid} symplectic state propagation scheme available in the MERCURY package \citep{mercury}. This scheme accurately models close encounters between test particles and planets by switching from mixed-variable symplectic to Bulirsch-Stoer state propagation within a certain distance from a planet, set to 2 Hill radii for this work. The solar system model included the eight major planets from Mercury to Neptune and was strictly Newtonian. Initial planetary state vectors were retrieved from the HORIZONS online ephemeris service \citep{Giorgini1996} at the J2000 epoch. Each asteroid was cloned 20 times with starting conditions for each clone generated by applying the linear transformation
\begin{equation}
\mathbf{y} = \mathbf{P} \mathbf{\Lambda}^{1/2} \mathbf{x} 
\end{equation}
to a 6-dimensional normally-distributed random variate $\mathbf{x}$ \citep{Duddy2012}. Here, $\mathbf{P}$ and $\mathbf{\Lambda}$ contain the  eigenvectors and eigenvalues of the asteroids’ state covariance matrices,  retrieved from the Near-Earth Objects Dynamic site for the epoch JD2459200.5 $=$ 2020 December 17.0 UT. Both massive bodies and test particles were integrated to the same epoch before the start of the simulations.

One question to be asked here is whether including the size-dependent Yarkovsky drag force in our dynamical model might change the outcome in a significant way. \citet{FenucciNovakovic2020} investigated this question for the Earth quasi-satellite (469219) Kamo'oalewa, an object comparable in both  size and orbit to the smallest object in our sample, 2017 SL16. Though Yarkovsky does change the orbital evolution of Kamo'oalewa over millions of yr and its residence time as an Earth co-orbital, actual differences from the gravity-only case were quite small and the overall effect on the evolution of the orbit was not significant. For this reason, and to minimise the computational overhead of our runs, we decided not to include the Yarkovsky effect in our simulations.

Two simulation batches were run for the four groups of asteroid clones plus the nominal orbits, one for $10^{4}$\,yr and the other for $10^{6}$\,yr, backwards and forwards from the starting epoch. The integration step size in both cases was 4\,days, the output step was $10$\,yr for the $10^{4}$\,yr runs and $10^{3}$\,yr for the $10^{6}$\,yr runs.

\begin{table*}[width=1.0\textwidth,cols=9,pos=t!]
\caption{Summary results of the numerical simulations. The four asteroids in our primary sample are indicated in bold.}\label{TBL:Sims}
\begin{minipage}{1.0\textwidth}\centering
\begin{tabular*}{\tblwidth}{@{} rlccrcccc@{} }
\toprule
 & &  &   & &   & \multicolumn{3}{c}{\hspace{1cm}1$-\sigma$ Dispersion\footnote{Dispersions for the five objects not in our primary sample are averages from the forward and backward runs}} \\
\multicolumn{2}{c}{Asteroid}                  & $a$   &      & \multicolumn{1}{c}{$I$}  &   Co-orbital & \multicolumn{2}{c}{for $a$ (au)/$e$/$I$ (deg)} & from Eq.~\ref{eq:disp} \\
(Number) & \multicolumn{1}{r}{Designation}                 & (au)  & $e$  & \multicolumn{1}{c}{(${}^{\circ}$)} & mode  & @$-$1Myr & @+1Myr & \\ \midrule
{\bf (138175)} &{\bf 2000 EE104}  & 1.004 & 0.29 &  5.2 &  N/A       & 0.24/0.14/4.3 & 0.30/0.14/4.1& 0.31 \\
{\bf (418849)} &{\bf 2008 WM64}  & 1.006 & 0.11 & 33.5 &  Passing   & 0.06/0.07/1.7 & 0.03/0.07/1.2 & 0.08\\
 &{\bf  2016 CA138}          & 0.999 & 0.05 & 27.7 &  Horseshoe & 0.18/0.13/4.6 & 0.07/0.10/3.7 & 0.17\\
 &{\bf 2017 SL16}            & 0.998 & 0.15 &  8.8 &  Quasi-Satellite/Horseshoe & 0.22/0.10/3.6 & 0.20/0.13/2.9 & 0.16\\
(468909) & 2014 KZ44         & 0.980 & 0.35 & 39.5   &  Passing   & \multicolumn{2}{c}{ 0.03/0.10/4} & 0.10\\
(468910) & 2014 KQ76         & 1.007 & 0.42 &  4.8   &  N/A       & \multicolumn{2}{c}{0.32/0.08/3} & 0.33\\
(512245) & 2016 AU8          & 0.982 & 0.20 &  9.2   &  Passing   & \multicolumn{2}{c}{0.20/0.07/5} &  0.21\\
(522684) & 2016 JP           & 0.994 & 0.38 & 11.3   &  N/A       & \multicolumn{2}{c}{0.15/0.09/6} & 0.17\\
& 2018 EB                    & 1.017 & 0.01 & 29.4   &  N/A       & \multicolumn{2}{c}{0.06/0.06/2} & 0.08 \\
\bottomrule
\end{tabular*}
\label{tab:stab}
\end{minipage}
\end{table*}

\subsection{Results}
Orbital evolution of the clone ensembles for each asteroid is presented in Figure~\ref{FIG:Simulations} where we also show the running mean and standard deviation for each clone ensemble. The standard deviation for the three actions $a$, $e$ and $I$  relative to the J2000 ecliptic  at $t=\pm 10^{6}$\,yr is separately reported in Table~\ref{tab:stab} and serves as a measure of orbital stability for each object. It is worth noting at this point that, according to our definition of stability, orbit transitions between coorbital modes while in the 1:1 resonance are considered stable since the semimajor axis continues to oscillate around 1\,au while $e$ and $I$ do not diffuse.  Below we discuss each asteroid separately.

\subsubsection*{\rm \bf (138175) 2000 EE104}
This asteroid has the most unstable orbit of those investigated in this work, with \mbox{$\sigma_{a}=0.05$\,au} after $\pm 10^{4}$\,yr and $\gtrsim$0.20\,au after $\pm 10^{6}$\,yr. This is probably due to its moderate eccentricity and low inclination, allowing frequent and relatively slow encounters with Venus as well as the Earth. None of the clones  remains within the Earth’s co-orbital region \mbox{($| a - 1\,\mbox{au} | < 0.01$\,au)} for more than a few hundred years from the start of the simulations. 

\subsubsection*{\rm \bf (418849) 2008 WM64} 
This asteroid is slowly drifting  backwards with respect to the Earth in what we refer to as a passing orbit \citep{Namouni1999}. Our simulations show that all orbits trace identical paths in $a$, $e$ and $I$ for at least $10^{4}$\,yr in the past and in the future. Longer-term, the asteroid has likely been in a passing orbit for the past 2$\times$10$^{5}$\,yr while the future evolution of the orbit is less certain, with the clone  semimajor axes beginning to disperse after a few times $10^{4}$\,yr. The orbit dispersion $10^{6}$\,yr from the present is $\sigma_{a}=0.06$\,au forwards and $0.03$\,au backwards, with five  clones remaining within the co-orbital region until $t=\pm 10^{6}$\,yr. Note the anti-correlated oscillations of $e$ and $I$, indicative of the Kozai regime with the argument of perihelion $\omega$ librating around $180^{\circ}$, typical of moderate-$I$ asteroids in the vicinity of the Earth’s orbit \citep{Michel1996,Namouni1999}. This state persists for at least 5$\times$10$^{5}$\,yr in the past and in the future and, together with the moderate-to-high inclination ($I\simeq34^{\circ}$), offers some protection against the orbit-changing effects of planetary encounters \citep{Michel1996}.

\subsubsection*{\rm \bf 2016 CA138}
This asteroid is currently in an Earth horseshoe orbit, qualifying therefore as the 13th Earth horseshoe \citep{Kaplan2020}.  Figure~\ref{FIG:angle} shows that the resonant angle $\Delta \lambda = \lambda - \lambda_{\rm Earth}$ for the clone orbits currently librates around $\Delta \lambda = 180^{\circ}$. All orbits enter the horseshoe phase $\sim$6$\times$10$^{3}$\,yr before $t=0$ and exit it $\sim$600\,yr after. We also note a short QS episode between $t = - 6.2 \times 10^{3}$ and $t = - 6.0\times 10^{3}$\,yr. In the forward runs, knowledge of the future orbital state begins to degrade after $\sim$10$^{3}$\,yr, however all orbits continue within some co-orbital mode. In our 1\,Myr runs, we find that confinement of the asteroid’s orbit within the Earth’s co-orbital region persists for several times $10^{4}$\,yr in the past and in the future. Interestingly, the final $a$ dispersion is higher in the backwards integrations (0.15\,au) compared to the forward runs (0.04\,au) with 1 clone still in the co-orbital region at $t=+1$\,Myr. Note also  the rapid increase of the eccentricity from initially $\sim$0.05 at $t=0$ to $0.20-0.30$ after $\sim$2$\times$10$^{5}$\,yr,  suggesting that the co-orbital resonance helps to keep $e$ below $\sim$0.15  in the few tens of thousands of years closest to $t=0$.

\subsubsection*{\rm \bf 2017 SL16}
\begin{center}
% \begin{sideways}
\begin{figure}
% \begin{minipage}{1.1\textwidth}
\centering
\includegraphics[width=\columnwidth]{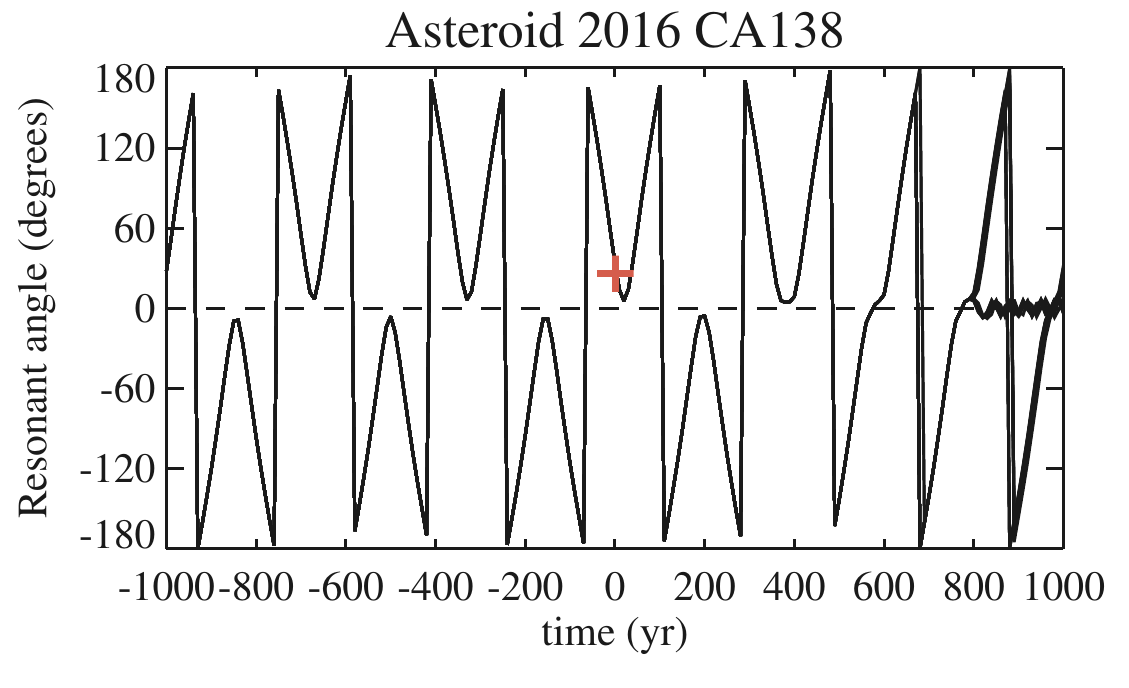}\\
\includegraphics[width=\columnwidth]{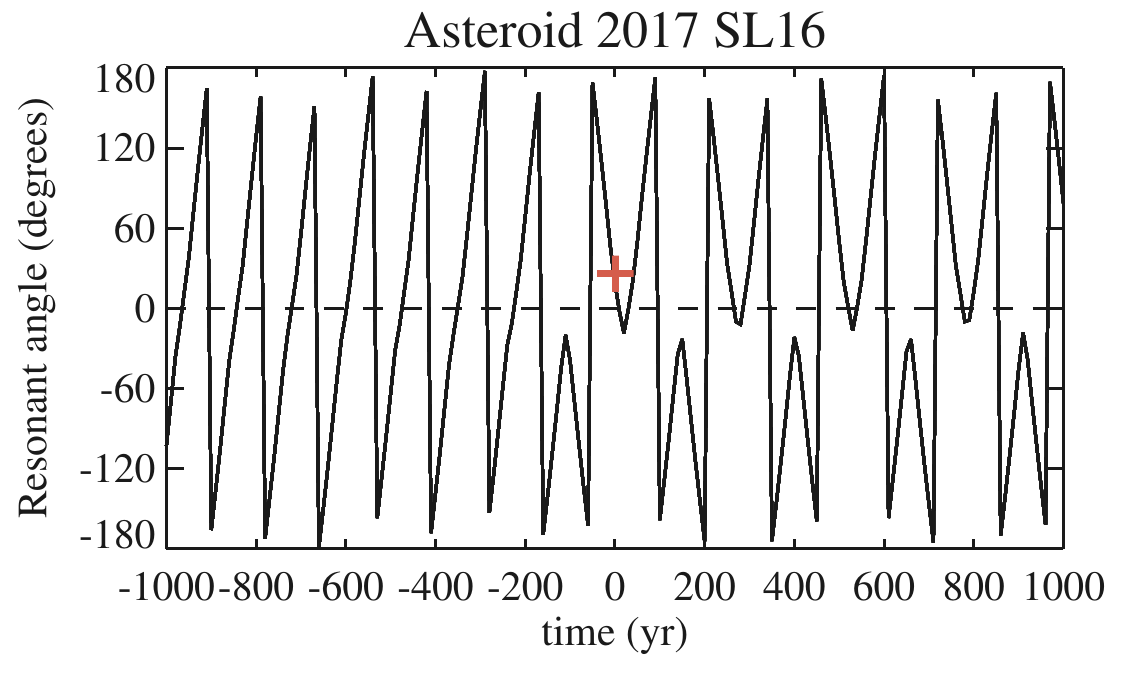}
\caption{ Evolution of the resonant angle $\Delta \lambda = \lambda - \lambda_{\rm Earth}$ for the clones of Earth co-orbitals 2016 CA138 (top) and 2017 SL16 (bottom) and for $10^{3}$ yr backwards and forwards from the present. The dashed horizontal line and red plus symbol indicate the $\Delta \lambda =0^{\circ}$ datum and the starting orbit respectively. }
\label{FIG:angle}
% \end{minipage}
% \end{sideways}
\end{figure}
\end{center}

The orbital evolution of this asteroid was recently investigated by \citet{Kaplan2020}. Those authors showed that SL16 is currently in an QS-HS asymmetric horseshoe configuration, having transitioned into this state from a passing orbit $\sim$100\,yr ago.  In addition to confirming the present QS-HS state (Fig.~\ref{FIG:angle}), our simulations of the asteroid’s orbital evolution up to $10^{4}$\,yr from the present are in very good agreement with \citet{Kaplan2020}. However, whereas those authors showed the asteroid to evolve to a higher $e$ and $I$ orbit in the future, our forward integrations show a different outcome, where $e$ and $I$ remain approximately constant from about $t=+3\times 10^{3}$\,yr until the end of the run. We attribute this to the improving orbit knowledge for this asteroid, with a better-determined orbit being available to us than in the \citeauthor{Kaplan2020} work. In the longer-term, the ensemble behaviour of $e$ and $I$ remains unchanged while the orbital semimajor axis disperses by as much as $0.2$\,au, with 3 of the 21 orbits still in the co-orbital region at $t=+1$\,Myr.

\subsection{Overall dynamical properties and relation to rotational state}
Here we combine the statistical dispersions for each orbital action in Table~\ref{tab:stab} into an ensemble indicator of orbital stability for each asteroid, defined as
\begin{equation}
\label{eq:disp}
\sigma^{2} = \left(\sigma_{a}/a\right)^{2} + \sigma_{e}^{2} + \left(\sigma_{I}/I\right)^{2}
\end{equation}

For this purpose, we carried out additional numerical simulations for the remaining five objects in the extended sample as for our original sample. None of these additional objects are currently locked in resonance, yet their orbital stability, as quantified by $\sigma$, are similar to those of the primary sample (Table~\ref{tab:stab}). The overall stability of these nine asteroids should therefore be representative of the population of NEAs with $a \sim$1\,au.

We find that the least stable objects in the extended sample are 138175 and 468910 ($\sigma > 0.3$) while the most stable ($\sigma < 0.1$) are 418849 and 2018 EB. The presence of an asteroid in the co-orbital resonance appears, therefore, to be a poor indicator of dynamical stability. Instead, the individual dispersions reported in Table~\ref{tab:stab} are correlated with the orbital actions, so that low $e$ and high $I$ orbits are the most stable.  

\begin{center}
% \begin{sideways}
\begin{figure*}
% \begin{minipage}{1.1\textwidth}
\centering
\includegraphics[width=0.9\columnwidth]{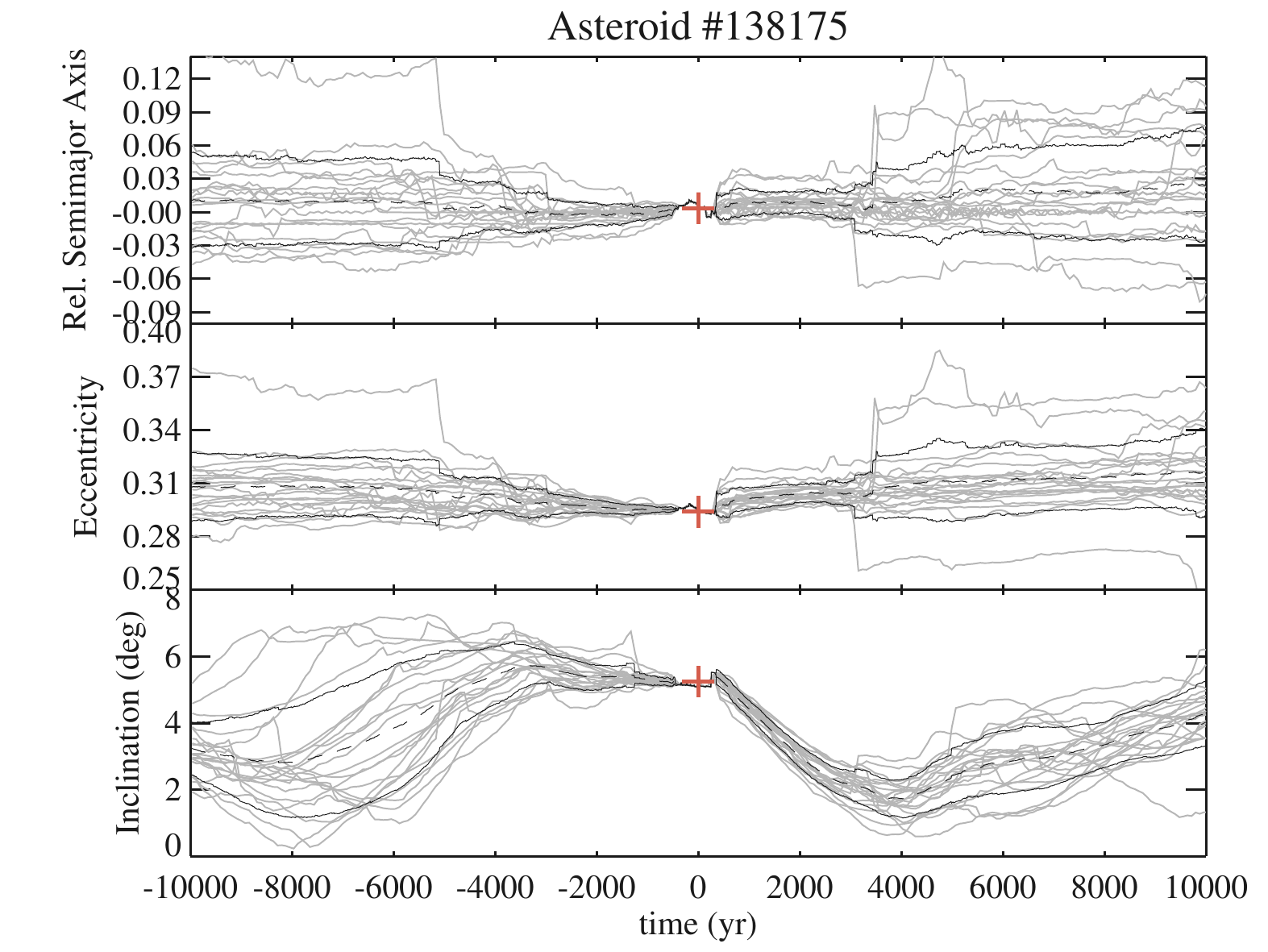}\includegraphics[width=0.9\columnwidth]{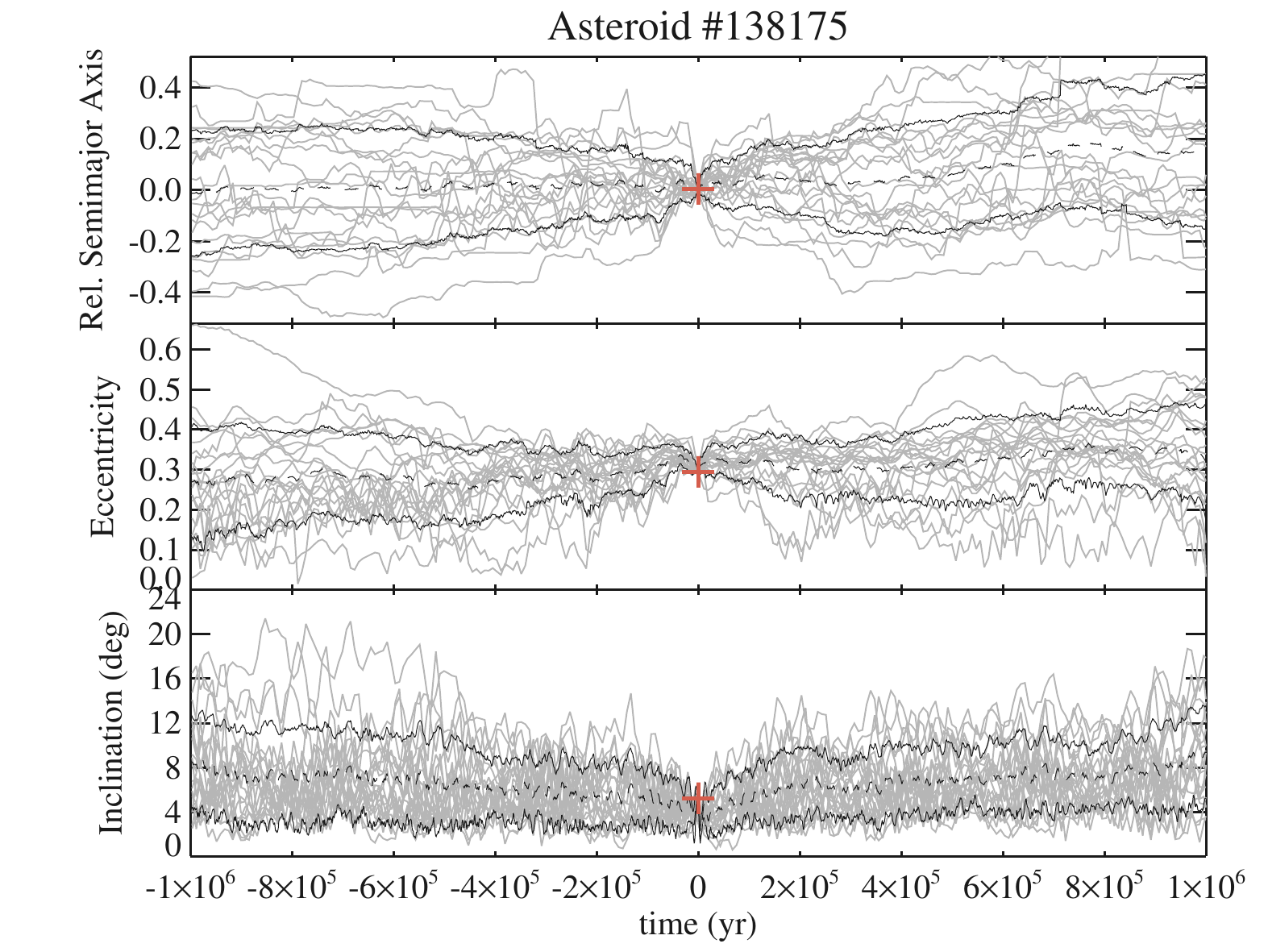}\\
\includegraphics[width=0.9\columnwidth]{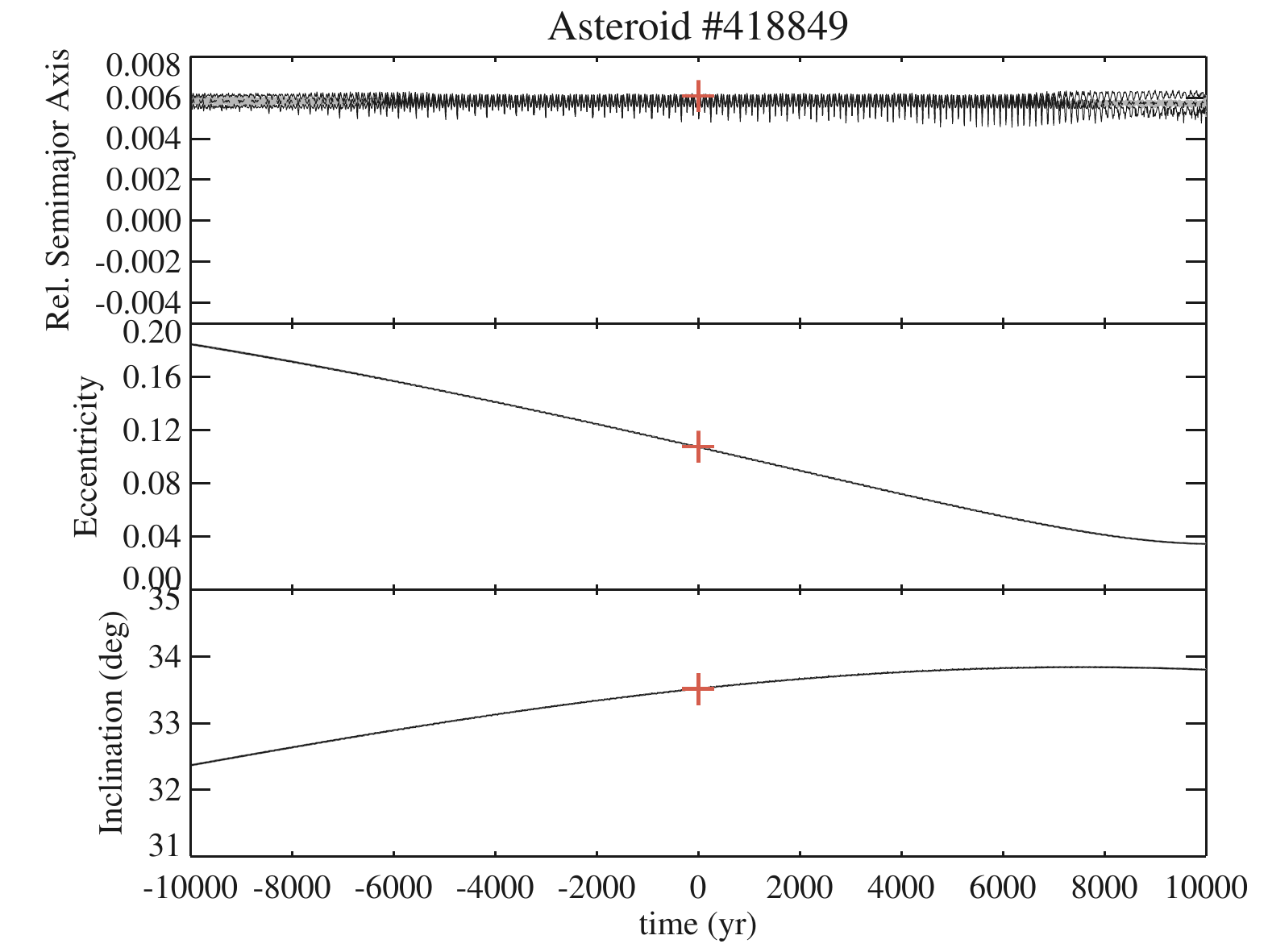}\includegraphics[width=0.9\columnwidth]{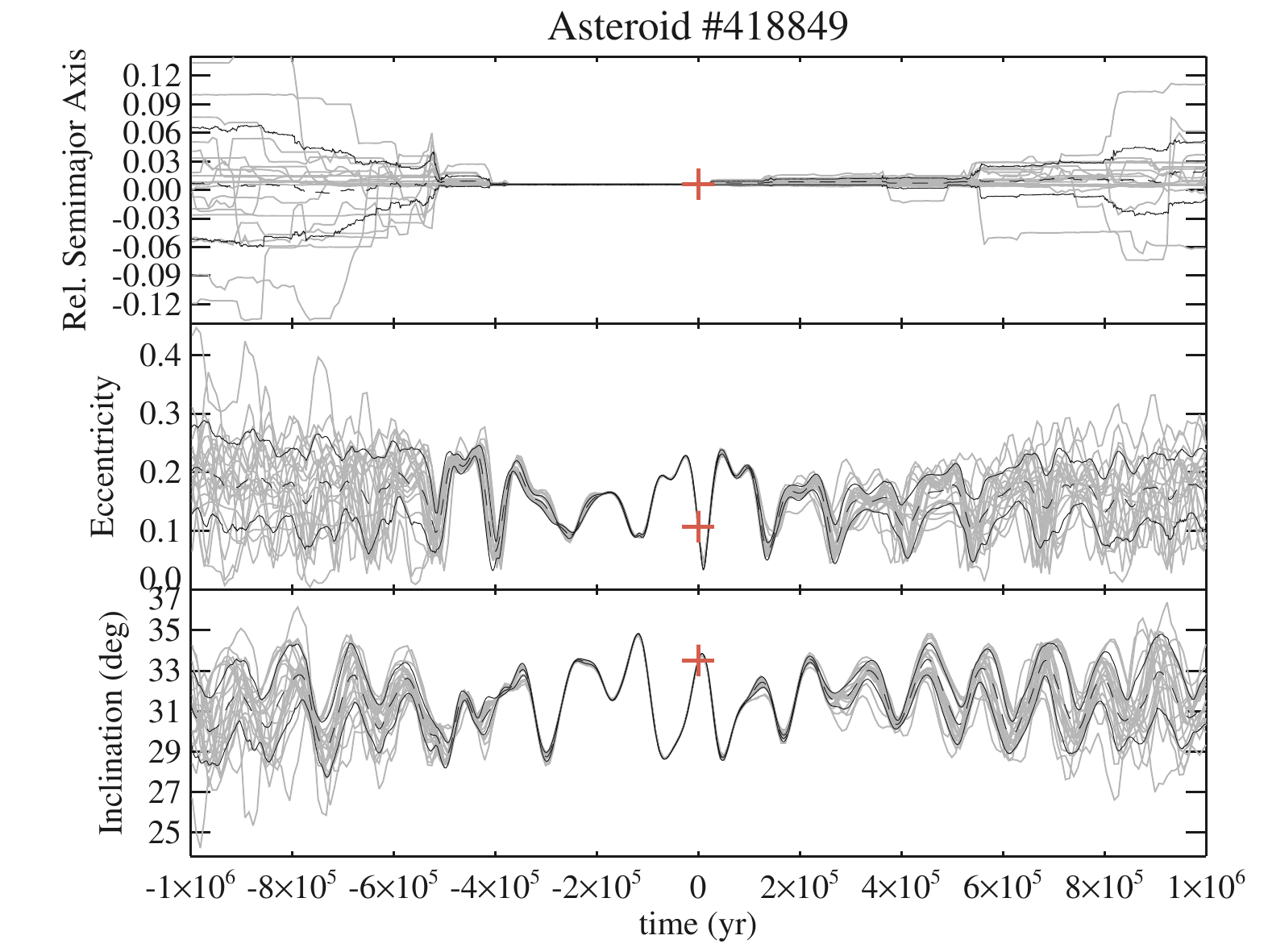}\\
\includegraphics[width=0.9\columnwidth]{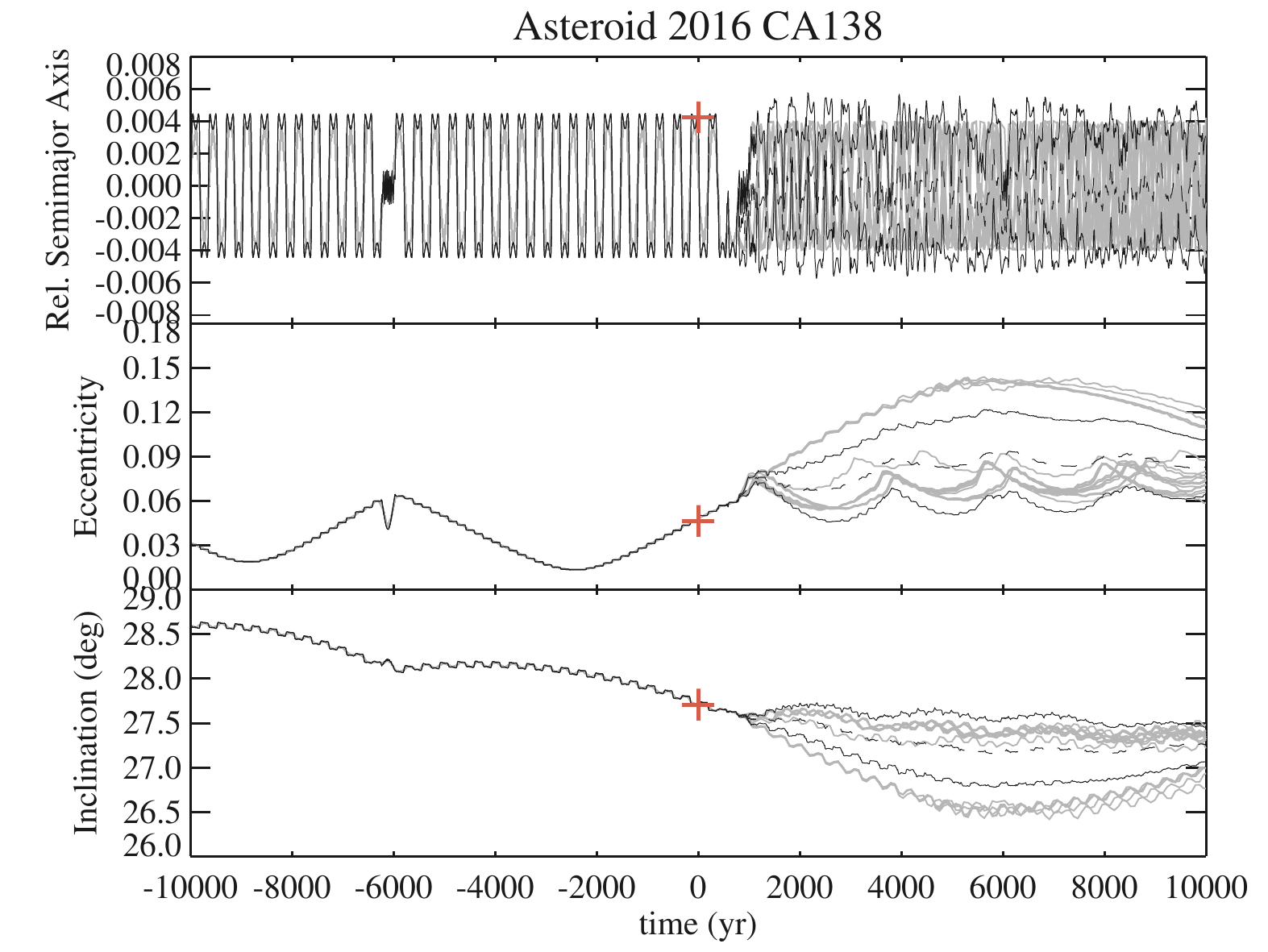}\includegraphics[width=0.9\columnwidth]{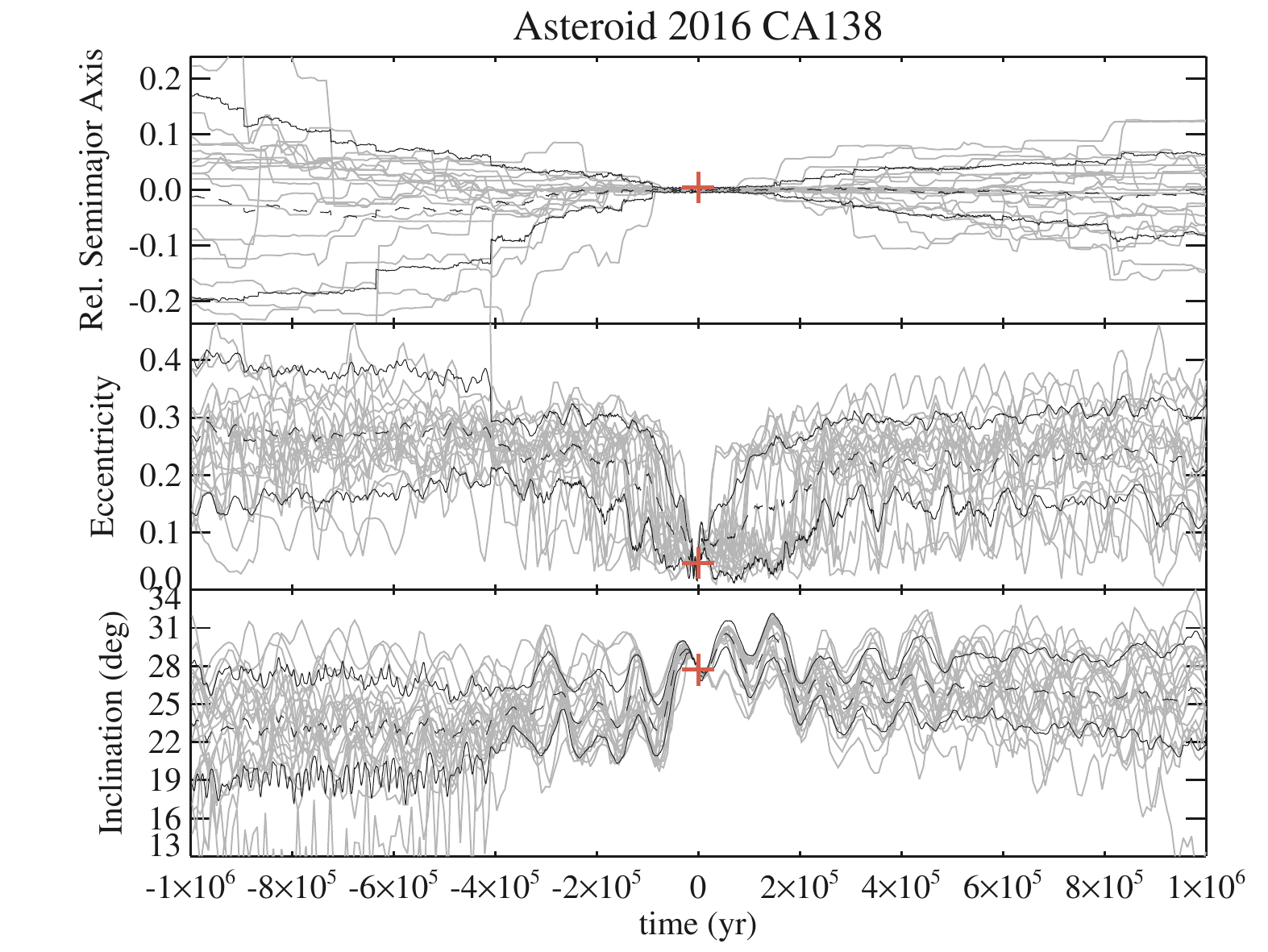}\\
\includegraphics[width=0.9\columnwidth]{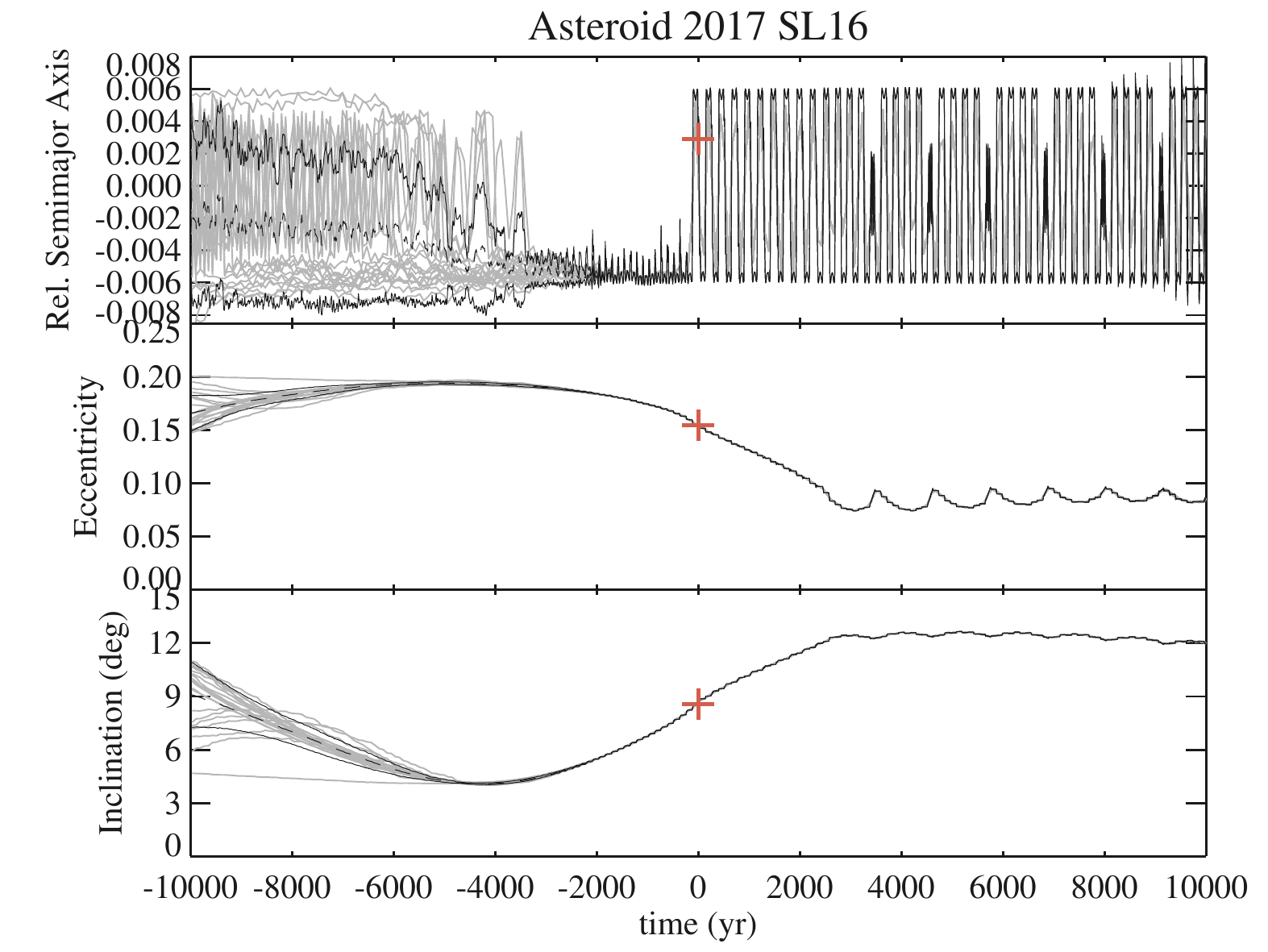}\includegraphics[width=0.9\columnwidth]{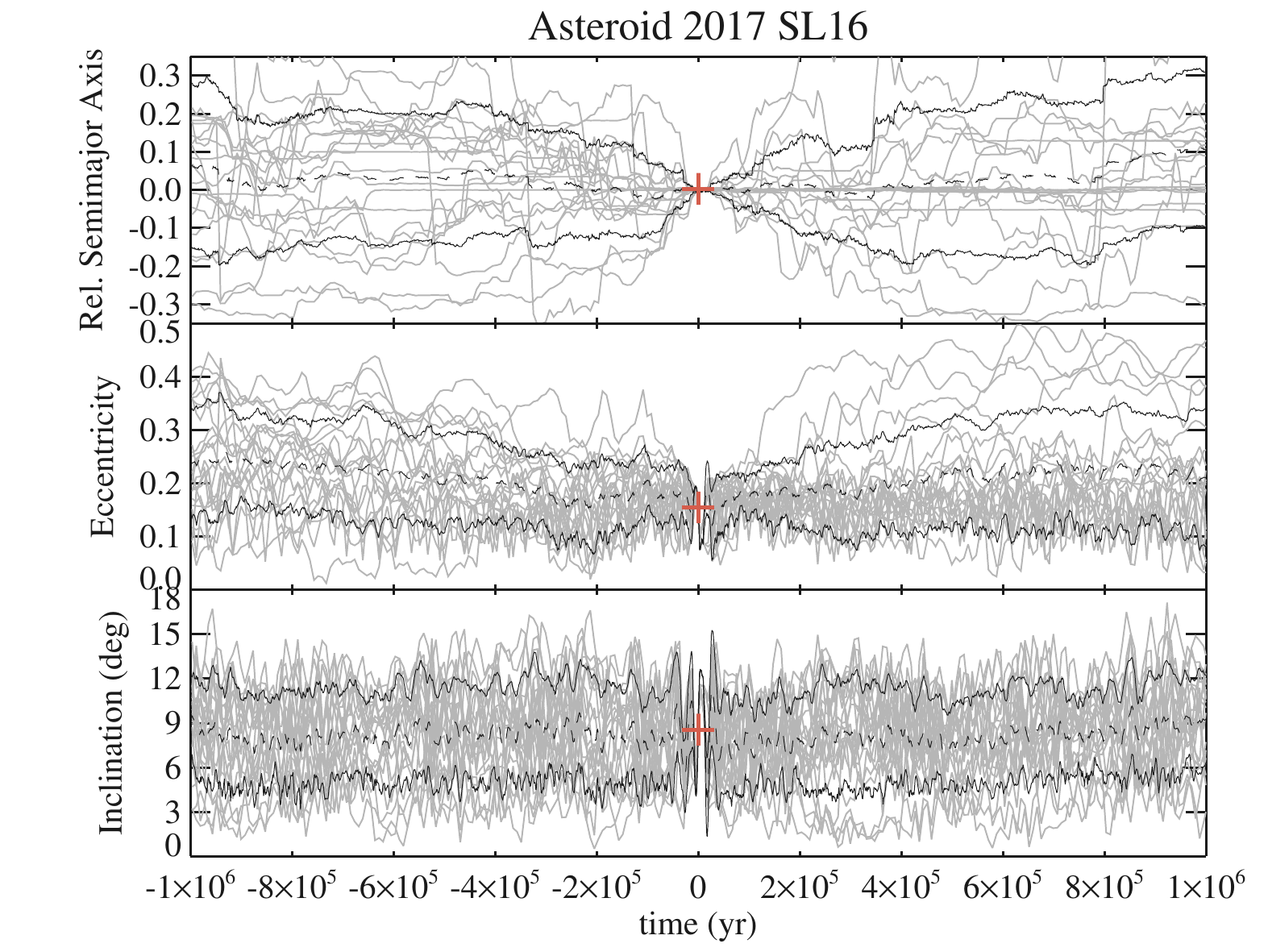}
\caption{Relative semimajor axis $(a - a_{\rm Earth})/a_{\rm Earth}$, eccentricity $e$ and inclination $I$  relative to the ecliptic plane  for the nominal orbit and 20 clones of each asteroid over $10^{4}$\,yr (left) and $10^{6}$\,yr (right) from $t=0$. Black dashed and solid lines represent the mean and standard deviation for each clone set. The red plus symbol indicates the starting orbit.}
\label{FIG:Simulations}
% \end{minipage}
% \end{sideways}
\end{figure*}
\end{center}

This suggests that the principal cause of instability is orbit-changing planetary encounters. Indeed, a high orbit inclination helps to avoid frequent close encounters with planets and moderates their orbit-changing effects even when they do occur \citep{Michel1996}. Furthermore, an asteroid in an orbit with $e\gtrsim0.25$ may approach Venus as well as the Earth. On the other hand, resonance-locking episodes and passing orbit states typically last $\sim$10$^4$\,yr (Figure~\ref{FIG:Simulations}, see also \cite{MoraisMorbidelli2002}) and much shorter than a Myr.

In this narrative, 138175 diffuses the fastest due to its high eccentricity allowing close encounters with Venus that disrupt the co-orbital resonance. In the same way, the moderate orbital diffusion displayed for 2016 CA138 and 2017 SL16 is caused by deterministic changes while in the 1:1 resonance with Earth and intermittent forays of the orbital eccentricity above the Venus-crossing value. Finally, the low-$e$, high-$I$ orbits of 418849 and 2018 EB mitigate against close approaches to both Earth and Venus. 

We return now to the question posed in the Introduction, namely whether orbital stability and the rotational state of NEAs at 1\,au are related. At first glance, this does not appear to be the case since, for example, the set of six asteroids near the spin rate barrier (Figure~\ref{FIG:Simulations}) is a mixture of stable and unstable cases (Table~\ref{tab:stab}).

Alternatively, we can choose to consider only those asteroids at the two extremes of the stability spectrum, ie those with the highest and lowest values of $\sigma$. If a relation of the type we are searching for existed, we would expect the rotational properties of those two groups to differ the most.

Here we find see that the spin rates of 418849, 468910 and 2018 EB are near-critical while that of 138175 is sub-critical. One might regard this as evidence that the variable dynamical environment of 138175 interferes with the spinning-up action of the YORP torque, preventing it from reaching a near-critical spin state. In a sample of only four, however, this evidence is far from conclusive and, while such a relationship might still exist, extending our study to a much larger sample of asteroids appears necessary to eke it out of the data.  

\section{Conclusions}
We determine the rotational periods and axis orientations  of four Earth co-orbital asteroids. 

The smallest object - 2017 SL16 with a diameter of a few tens of meters - has a rotational period P=0.3188\,hrs which suggests that it is monolithic. 
Even though we cannot find a definitive pole solution, we can say that it is likely not near the ecliptic and the best solution  from the five models with similar $\chi^2$ values has coordinates  \mbox{$\lambda_1$=190.44$^{\circ}$} and \mbox{$\beta_1$=34.32$^{\circ}$}

The other extreme case is the asteroid (138175) 2000 EE104, which we find to be a slow rotator with a spin rate of P=13.9476\,hrs. The pole position is also not uniquely determined but most probably it lies above the ecliptic with the formal best-fit pole at \mbox{$\lambda_1$=233.86$^{\circ}$} and \mbox{$\beta_1$=68.71$^{\circ}$}.

The object with the best quality of observations and respectively the best results is (418849) 2008 WM64. It has only one clear definitive solution for the rotational period: P=2.4077\,hrs. Also, the pole solution lies on a single spot of the $\chi^2$ search map at coordinates \mbox{$\lambda_1$=217.34$^{\circ}$} and \mbox{$\beta_1$=-39.17$^{\circ}$}.

For the second smallest object in our sample - 2016 CA138 with an estimated diameter of 75\,m - we determine the spin rate to be 5.3137\,hrs. The $\chi^2$ map shows two vertical stripes with low $\chi^2$ values at longitudes 180$^\circ$ apart - around 80$^\circ$ and 260$^\circ$.

We compared the rotational properties and sizes of our asteroid sample to NEA entries in the LCDB database \citep{lcdb_v4_2021}. Our overall conclusion is that the size vs rotational frequency distribution of the co-orbital asteroids appears not to differ from the one of the NEAs.

We made numerical simulations in order to investigate if there is a relation between the orbit stability and the rotational state of Earth co-orbitals. Our results show that the co-orbital resonance is not affecting the orbit stability, but the orbit itself is responsible for that and mainly its eccentricity ($e$) and inclination ($I$). Orbits with low $e$ and high $I$ are the most stable because firstly  highly-inclined orbits are relatively stable against  close encounters with planets and secondly orbits with high $e$ may approach Venus as well as the Earth.

We cannot make a definitive conclusion if the orbit stability and rotational state of the asteroids are related, so we need further investigations and observations to increase our sample in order to obtain more statistically significant results.

\section{Acknowledgements}
This work was supported via grant ST/R000573/1 from the UK Science and Technology Facilities Council. The authors gratefully acknowledge observing grant support from the Institute of Astronomy and National Astronomical Observatory, Bulgarian Academy of Sciences. Astronomical research at the Armagh Observatory \& Planetarium is grant-aided by the Northern Ireland Department for Communities (DfC). The authors also acknowledge DfC for the FoReRo2 instrument development contribution from the Armagh Observatory \& Planetarium toward the new CCD camera Andor iKon-L used in this study.

%\mbox{} 
%\vspace{-0.8cm}

%\mbox{} 

\bibliographystyle{cas-model2-names}
\bibliography{CoOrbs}
\end{document}